\documentclass[3p]{elsarticle}

\usepackage{lineno,hyperref}
\usepackage{amssymb}
\usepackage{amsmath}
\usepackage {graphicx}
\usepackage{array}
\usepackage{tikz}
\usetikzlibrary{decorations.markings}
\usepackage{soul}
\usepackage{caption}
\usepackage{hyperref}
\usepackage{subcaption}
\usepackage{booktabs}
\usepackage{xcolor}
\usepackage{tabularx}
\usepackage{setspace}

\journal{Physics of Plasmas}

\newcommand{\norm}[1]{\left\lVert#1\right\rVert}

\newcommand{\refeq}[1]{Eq.(\ref{#1})}
\newcommand{\refeqs}[2]{Eqs.(\ref{#1})--(\ref{#2})}

\newcommand{\refequation}[1]{Equation (\ref{#1})}

\newcommand{\reffig}[1]{Fig.(\ref{#1})}
\newcommand{\reffigure}[1]{Figure (\ref{#1})}

\newcommand{\Sp}[1]{{\cal #1}}

\newcommand{\sO}{\Sp{O}}
\newcommand{\sP}{\Sp{P}}

\newcommand{\V}[1]{{\bf #1}}

\newcommand{\vb}{\V{b}}

\newcommand{\vj}{\V{j}}

\newcommand{\vr}{\V{r}}

\newcommand{\vv}{\V{v}}

\newcommand{\vx}{\V{x}}
\newcommand{\vy}{\V{y}}

\newcommand{\vB}{\V{B}}

\newcommand{\vE}{\V{E}}
\newcommand{\vF}{\V{F}}
\newcommand{\vG}{\V{G}}

\newcommand{\vX}{\V{X}}

\newcommand{\vzeta}{\boldsymbol{\zeta}}

\newcommand{\M}[1]{{\mathbb #1}}
\newcommand{\mA}{\M{A}}

\newcommand{\mI}{\M{I}}

\newcommand{\mM}{\M{M}}

\newcommand{\mW}{\M{W}}

\newcommand{\pd}[2]{\frac{\partial #1}{\partial #2}}
\newcommand{\ppd}[2]{\frac{\partial^2 #1}{\partial #2^2}}

\newcommand{\h}{\hat}

\newcommand{\pll}{\parallel}

\begin{document}

\begin{frontmatter}

\title{Verification of a Fully Implicit Particle-in-Cell Method for the \texorpdfstring{$v_\pll$}{Lg} Formalism of Electromagnetic Gyrokinetics in the XGC Code}

\author[address1]{Benjamin J. Sturdevant \corref{mycorrespondingauthor}}
\ead{bsturdev@pppl.gov}
\author[address1]{S. Ku}
\author[address2]{L. Chac\'{o}n}
\author[address4]{Y. Chen}
\author[address5]{D. Hatch}
\author[address1]{M. D. J. Cole}
\author[address1]{A. Y. Sharma}
\author[address3]{M. F. Adams}
\author[address1]{C. S. Chang}
\author[address4]{S. E. Parker}
\author[address1]{R. Hager}

\address[address1]{Princeton Plasma Physics Laboratory, P.O. Box 451, Princeton, NJ 08543, USA}
\address[address2]{Los Alamos National Laboratory, P.O. Box 1663, Los Alamos, NM 87545, USA}
\address[address3]{Lawrence Berkeley National Laboratory, 1 Cyclotron Rd., Berkeley, CA 94720, USA}
\address[address4]{University of Colorado at Boulder, Department of Physics, 390 UCB, Boulder, CO 80309, USA}
\address[address5]{Institute for Fusion Studies, University of Texas at Austin, 1 University Station, C1500, Austin, TX 78712}
\cortext[mycorrespondingauthor]{Corresponding author}

\begin{abstract}
A fully implicit particle-in-cell method for handling the $v_\pll$-formalism of electromagnetic gyrokinetics has been implemented in XGC. By choosing the $v_\pll$ formalism, we avoid introducing the non-physical skin terms in Amp\`{e}re's law, which are responsible for the well-known ``cancellation problem" in the $p_\pll$-formalism. The $v_\pll$-formalism, however, is known to suffer from a numerical instability when explicit time integration schemes are used due to the appearance of a time derivative in the particle equations of motion from the inductive component of the electric field. Here, using the conventional $\delta f$ scheme, we demonstrate that our implicitly discretized algorithm can provide numerically stable simulation results with accurate dispersive properties. We verify the algorithm using a test case for shear Alfv\'{e}n wave propagation in addition to a case demonstrating the ITG-KBM transition. The ITG-KBM transition case is compared to results obtained from other $\delta f$ gyrokinetic codes/schemes, whose verification has already been archived in the literature.
\end{abstract}


\end{frontmatter}


\section{Introduction}

Gyrokinetic particle-in-cell codes are used extensively to study instabilities and microturbulence in magnetically confined fusion devices. Electrostatic gyrokinetic particle simulations of ion temperature gradient (ITG) driven microturbulence have been accessible now for a few decades with the adiabatic electron model \cite{Parker1993PRL,Dimits1996,Sydora1996,Lin1998}. More challenging is the kinetic treatment of electrons due to their small mass and the presence of additional high-frequency modes \cite{Lee1987}. Specialized numerical techniques are generally used to mitigate these issues \cite{Ku2016,Ku2018,Manuilskiy2000,Chen2001,Chen2003,Chen2007,Adams1982}. There are important physics effects in magnetically confined fusion devices that, in addition to kinetic electrons, require the inclusion of electromagnetic perturbations. Electromagnetic capabilities have been implemented in a number of particle and continuum gyrokinetic codes including GEM \cite{Chen2001,Chen2003,Chen2007}, GTC \cite{Lin1998,Bao2017,Bao2018,Dong2019}, EUTERPE \cite{Kornilov2004,Kleiber2016}, ORB5 \cite{Jolliet2007,Bottino2011,Lanti2020}, GENE \cite{Jenko2000,Gorler2011,Gorler2011PoP}, Gkeyll \cite{Mandell2020,Hakim2020}, GYRO \cite{Candy2003,Candy2005}, GKV \cite{Watanabe2006,Maeyama2014}, and GKNET \cite{Obrejan2017,Ishizawa2019}. However, application to long-wavelength MHD-type modes is often limited due to numerical difficulties.
\par For particle codes in particular, handling the parallel vector potential $A_\pll$ at high $\beta$ (ratio of plasma to magnetic pressure), long perpendicular wavelength regimes remains a challenge. It is common to use the parallel canonical momentum $p_\pll$ as a coordinate, which leads to the appearance of a large skin current term in Amp\`{e}re's law from the zero-order distribution \cite{Chen2001,Cummings1994}. To avoid severe accuracy problems, this skin current term needs to cancel with the corresponding part of the current deposited from the marker particles. The difference in numerical representations between the current appearing in Amp\`{e}re's law and the current coming from marker particles makes this exact cancellation particularly difficult to achieve. When the parallel component of velocity $v_\pll$ is instead used as a coordinate, the original form of Amp\`{e}re's law can be used without the appearance of skin current terms. Hence, the $v_\pll$ formalism avoids the cancellation problem altogether. The difficulty with this approach, however, is the appearance of a time derivative in the particle equations of motion from the inductive component of the electric field perturbation, $\pd{A_\pll}{t}$. Explicit time integration methods are generally unstable with the inclusion of this term \cite{Reynders1993}.
\par Kinetic electron capabilities in XGC \cite{Ku2016,Ku2018,Chang2009} are already well established and have been used in large scale physics studies of the tokamak edge \cite{Chang2017,Chang2017PRL,HagerPoP2019,Chang2021}. There have been recent efforts to extend this capability to include electromagnetic perturbations. Besides the approach described in this paper, an explicit method using the mixed variables/pullback transformation scheme \cite{Mishchenko2014a,Mishchenko2014b,Kleiber2016} has been recently implemented \cite{Cole2021}. In addition, an implicit kinetic-fluid hybrid model has been previously explored as an inexpensive alternative to electron particles \cite{Hager2017}. A key goal of these efforts is to develop the capability of simulating microturbulence consistently with MHD-type electromagnetic modes in the full volume of magnetically confined fusion devices from the magnetic axis to the first wall. Such a capability has not been previously demonstrated. However, we mention that there have been notable steps taken toward this end. For example, ORB5 has recently achieved simulations capturing the self-consistent interactions between Alfv\'en modes and ITG turbulence using a global gyrokinetic model in a simple core geometry \cite{Biancalani2021}. Beyond MHD, an electromagnetic version of XGC will be useful for the study of microtearing modes and electromagnetic effects on electrostatic modes, e.g., finite-beta stabilization. A full-volume electromagnetic capability would represent an important step forward both for XGC and for gyrokinetic particle simulations in general.
\par Here, we have applied the techniques developed in Refs. \cite{GChen2011,Chacon2013,GChen2014,GChen2014JCP,GChen2015} in the 6-dimensional (6D) particle-in-cell context to enable a fully implicit particle-in-cell method for handling the $v_\pll$-formalism of electromagnetic gyrokinetics in the 5D particle-in-cell code XGC. Besides the work we document in this paper on XGC, we mention recent efforts in developing an implicit gyrokinetic electromagnetic scheme in the mixed particle-in-cell/particle-in-Fourier TRIMEG-GKX code for applications to energetic particle physics \cite{Lu2021}. Implicit time discretization is effective for eliminating the stability issues originating from the inductive component of the electric field, and by using the $v_\pll$-formalism, we avoid the cancellation problem in Amp\`{e}re's law. The implicit approach, however, requires the solution of a large system of nonlinear equations at each time step, for which we employ Anderson mixing \cite{Anderson1965} to a preconditioned Picard iteration scheme. The focus of this paper is to present the 5D equations and algorithms, and to verify the scheme implemented in XGC for two test cases. The first demonstrates shear Alfv\'{e}n wave (SAW) propagation in cylindrical geometry in the long perpendicular wavelength regime, and the second demonstrates the transition from the ITG instability to the kinetic ballooning mode (KBM) in toroidal geometry as $\beta$ is increased past a critical value. These test cases were motivated by ones previously considered in \cite{Ma2018} for the GTS code \cite{Wang2006,Wang2010}. By comparing to an analytical dispersion relation, the implicit scheme is shown to accurately reproduce the dispersion properties of the SAW in regimes inaccessible with the $p_\parallel$-formalism. In addition, the implicit scheme shows good agreement with the GEM and GENE codes and the explicit electromagnetic implementation in XGC \cite{Cole2021} for the ITG-KBM transition case. The results presented in this paper strengthen our confidence in the ability of the implicit scheme to accurately solve the electromagnetic gyrokinetic equations. 
\par The remainder of the paper is organized as follows. In Sec.~\ref{sec:model}, we present the equations implemented in XGC based on the $v_\pll$-formalism of electromagnetic gyrokinetics. In Sec.~\ref{sec:implicit}, we give an overview of the implicit algorithm, including the discretization of the equations and our approach to solve the resulting system of equations at each timestep. In Sec.~\ref{sec:verification}, we describe the two test cases used to verify our implementation and present the results. In Sec.~\ref{sec:performance}, we briefly discuss the performance of the implicit algorithm. Finally, we give conclusions in Sec.~\ref{sec:conclusions}.


\section{Gyrokineitc Electromagnetic Model in the $v_\pll$ Formalism} \label{sec:model}
The $v_\pll$-formalism of the electromagnetic gyrokinetic equations requires the parallel component of the perturbed vector potential $A_\pll$, which is obtained from the following form of Amp\`{e}re's law:
\begin{align}
-\nabla_\perp^2 A_\pll = \mu_0 \sum_{s=i,e} \langle \delta j_{\pll s} \rangle, \label{eq:ampere}
\end{align}
where $\mu_0$ is the permeability of free space, and $\langle \cdot \rangle$ represents the gyroaveraging operator. Throughout, it is understood that gyroaveraging is only performed for the ions, as the electrons are treated as drift-kinetic.  This simplification can be easily removed. The perturbed current contribution from each species on the right hand side is deposited from the marker particles using the $\delta f$ weights \cite{Parker1993,Dimits1993,Hu1994,Aydemir1994}. The background current is balanced out by the equilibrium magnetic field and plasma. The gyrokinetic Poisson equation is used to solve for the electrostatic potential $\phi$ and is given by:
\begin{align}
- \nabla \cdot \frac{m_i n_{0i}}{B^2} \nabla_\perp \phi = \sum_{s=i,e} q_s \langle \delta n_s \rangle, \label{eq:gkpoisson}
\end{align}
where $m_i$ is the ion mass, $n_{0i}$ is the background ion density, $B$ is the background magnetic field strength, and $q_s$ is the charge for species $s$. The short wavelength correction term often used in XGC \cite{Ku2018} is turned off in this initial study for simplicity, which has been justified in the benchmarking paper Ref. \cite{Cole2021}. Again, the perturbed number densities $\delta n_s$ on the right hand side come from marker particles, and gyroaveraging is assumed only for the ions. From $\phi$ and $A_\pll$, we can compute the perturbed electric and magnetic fields as
\begin{align}
\delta \vE &= - \nabla \phi - \pd{A_\pll}{t} \h{\vb} \\
\delta \vB &= \nabla A_\pll \times \h{\vb}, \label{eq:deltaB}
\end{align}
where $\h{\vb}$ is the direction of the background magnetic field. Here, we note the presence of the time derivative in the second term of the electric field, which leads to numerical instabilities when explicit time integration schemes are used. Here, we do not compute $\delta \vB$ from the curl of $A_\pll \h{\vb}$ as is done, for example, in Ref. \cite{Hatzky2019}. The difference represents a higher order corrction in the gyokinetic ordering, which we have neglected in the present study. The particle equations of motion for species $s = i,e$ can be written as :
\begin{align}
\dot{\vX} &= v_\pll \frac{\vB^*}{B_\pll^*} + \left( \langle \delta \vE \rangle - \frac{\mu}{q_s} \nabla B \right) \times \frac{\h{\vb}}{B_\pll^*} \label{eq:xdot} \\ 
\dot{v}_\pll &= \frac{q_s}{m_s} \frac{\vB^*}{B_\pll^*} \cdot \left( \langle \delta \vE \rangle - \frac{\mu}{q_s} \nabla B \right), \label{eq:vdot}
\end{align}
where $\vX$ is the gyrocenter position vector, $v_\pll$ is the parallel component of velocity, $\mu$ is the magnetic moment, $m_s$ is the mass of species $s$. We note that, in this form of the gyrocenter equations of motion, the curvature drift is hidden in the first term $\vB^* / B_\pll^*$, with
\begin{align}
\vB^* &= \vB + \langle \delta \vB \rangle + \frac{m_s}{q_s} v_\pll \nabla \times \h{\vb} \\
B_\pll^* &= \h{\vb} \cdot \vB^*.
\end{align}
For the purposes of this report, we use the conventional $\delta f$ method, splitting the distribution functions for each species into background and perturbed parts as $f_s = f_{0 s} + \delta f_s$. This splitting leads to an equation for the evolution of particles weights as :
\begin{align}
\frac{d w}{dt} = (1-w) \delta  \dot{f}_s / f_{0 s},
\end{align}
where
\begin{align}
\delta \dot{f}_s = -\dot{\vX}_1 \cdot \nabla f_{0s} - \dot{v}_{\pll 1} \pd{f_{0 s}}{v_\pll}. \label{eq:dfdot}
\end{align}
Here, the subscript $1$ indicates that only perturbed quantities are kept from \refeqs{eq:xdot}{eq:vdot}. In particular,
\begin{align}
\dot{\vX}_1 &= v_\pll \frac{\langle \delta \vB \rangle}{B_\pll^*} + \langle \delta \vE \rangle  \times \frac{\h{\vb}}{B_\pll^*} \label{eq:x1dot} \\ 
\dot{v}_{\pll 1} &= \frac{q_s}{m_s} \cdot \left( \frac{\vB^*}{B_\pll^*} \cdot \langle \delta \vE \rangle - \frac{\mu}{q_s} \frac{\langle \delta \vB \rangle}{B_\pll^*} \cdot \nabla B \right) \label{eq:v1dot}.
\end{align}
We note that $B^*_\pll$ does not contain any perturbed quantities, since $\langle \delta \vB \rangle$ computed from \refeq{eq:deltaB} is perpendicular to $\h{\vb}$. If $\delta \vB$ were computed using the full curl of $A_\pll \h{\vb}$, rather than from \refeq{eq:deltaB}, a contribution from $A_\pll$ would appear in $B_\pll^*$ as in Ref. \cite{Hatzky2019}.

\par Finally, we mention that \refeq{eq:dfdot} is only consistent with the gyrokinetic Vlasov equation when $f_{0s}$ is an exact equilibrium solution of the gyrokinetic Vlasov equation. For a local Maxwellian $f_{0s}$, contributions due to the grad-B and curvature drifts drivers can be missed in the weight equation using \refeq{eq:dfdot}. These contributions are commonly ignored in conventional $\delta f$ codes and are ignored in the code used in this report for cross-verification in Sec.~\ref{sec:verification}.


\section{Implicit Algorithm} \label{sec:implicit}

In this section, we describe the implicit time discretization scheme applied to the system in Sec.~\ref{sec:model}, as well as the iterative scheme and preconditioner used in solving the resulting nonlinear system of equations. 
The discretization scheme and iterative solution method are based on the work in \cite{GChen2011,Chacon2013,GChen2014,GChen2014JCP,GChen2015}. 

\subsection{Time Discretization} \label{sec:implicit_timedisc}
In our scheme, electrons are subcycled \cite{Adams1982,Cohen1985,GChen2011,Sturdevant2016b}, meaning they are advanced using several small time steps over the interval $n \Delta t \le t \le (n+1) \Delta t$. Hence, we need to define the perturbed electric and magnetic fields over the continuous time interval between steps $n$ and $n+1$. Following \cite{GChen2011,Chacon2013,GChen2014,GChen2014JCP,GChen2015}, we take the electric field to be constant in time over the subcycling interval as
\begin{align}
\delta \vE(t) = - \nabla \left( \frac{\phi^{n+1} + \phi^n}{2} \right) - \frac{2}{\Delta t} \left( A_\pll^{n+1/2} - \tilde{A}_\pll^n \right) \h{\vb} \ \ \ \mathrm{for} \ \ n \Delta t \le t \le (n+1) \Delta t, \label{eq:deltaEoft}
\end{align}
where $\tilde{A}_\pll^n$ is defined recursively in time by
\begin{align}
\tilde{A}_\pll^n = 2 A_\pll^{n-1/2} - \tilde{A}_\pll^{n-1},
\end{align}
and is initialized by solving \refeq{eq:ampere} at timestep $0$. Consistent with the time derivative of the parallel vector potential being constant within the interval, we take $\delta \vB$ to vary linearly in time as \cite{GChen2014,GChen2014JCP,GChen2015}:
\begin{align}
\delta \vB(t) = \left[ \left(1 - \frac{t- n \Delta t}{\Delta t/2} \right) \nabla \tilde{A}_\pll^n +  \left( \frac{t- n \Delta t}{\Delta t/2} \right) \nabla A_\pll^{n+1/2} \right] \times \h{\vb}. \label{eq:deltaBoft}
\end{align}
The gyrocenter positions of marker particles are described using a cylindrical coordinate system $(R,Z,\varphi)$, where $R$ is the major radius, $Z$ is the distance along the cylindrical axis, and $\varphi$ is the toroidal angle. Together with $v_\pll$ and the $\delta f$ particle weight $w$, there are five evolving variables for each marker particle. These can be written in vector form as $\vzeta = [R,Z,\varphi,v_\pll,w]^T$, and marker particle evolution can then be expressed as 
\begin{align}
\dot{\vzeta} = \vG(\vzeta,t), \label{eq:particleODE}
\end{align}
where
\begin{equation}
\arraycolsep=8pt\def\arraystretch{2.2}
\vG(\vzeta,t) = \left[
\begin{array}{c}
\left( v_\pll B^*_R + F_Z \frac{B_\varphi}{B} - F_\varphi \frac{B_Z}{B} \right) / B^*_\pll \\
\left( v_\pll B^*_Z + F_\varphi \frac{B_R}{B} - F_R \frac{B_\varphi}{B} \right) / B^*_\pll \\
\left( v_\pll B^*_\varphi + F_R \frac{B_Z}{B} - F_Z \frac{B_R}{B} \right) / R B^*_\pll \\
\frac{q_s}{m_s} \left( B^*_R F_R + B^*_Z F_Z + B^*_\varphi F_\varphi \right) / B^*_\pll \\
-(1-w) \left( \dot{\vX}_1 \cdot \nabla \Psi \pd{}{\Psi} \ln{f_{0s}} + \dot{v}_{\pll 1} \pd{}{v_\pll} \ln{f_{0s}} \right)
\end{array}
\right]. \label{eq:rhsG}
\end{equation}
In this expression, we have defined the vector quantity $\vF = \langle \delta \vE \rangle - \frac{\mu}{q_s} \nabla B$ and have assumed spatial variations in $f_{0s}$ to be along the poloidal flux coordinate $\Psi$. Given the discrete-time potentials $\phi^{n}$, $\phi^{n+1}$, $\tilde{A}_\pll^n$, and $A_\pll^{n+1/2}$ defined on the mesh, \refeq{eq:deltaEoft} and \refeq{eq:deltaBoft} together with the particle interpolation methods allow for the evaluation of the right hand side vector $\vG(\vzeta,t)$ at all particle locations $\vzeta$ within the mesh and all times within the interval $n \Delta t \le t \le (n+1) \Delta t$. This allows considerable freedom in choosing an ODE integration method to advance \refeq{eq:particleODE} in time from step $n$ to $n+1$. Here, we choose a standard fourth order Runge-Kutta method (RK4) for both electrons and ions. For subcycled electrons, we divide the interval into an integer number of sub-intervals and apply RK4 successively over the sub-intervals to advance from step $n$ to $n+1$. Adaptive integration schemes such as the one considered in Ref. \cite{GChen2011} are well-suited for subcycling and will be explored in the future. Ions are not subcycled and use field values at the center and ends of the time interval to compute the RK4 stages.

\subsection{Residual Evaluation}

By choosing an implicit time discretization, the particles and fields are interdependent at each timestep. Particles are pushed using fields that depend on $\phi^{n+1}$ and $A_\pll^{n+1/2}$, yet these potentials are determined from the particle states over the interval $n \Delta t \le t \le (n+1) \Delta t$. In particular, the right hand side of Amp\`{e}re's law is computed from the time averaged current depositions over the subcycled timesteps between $n$ and $n+1$, and the right hand side of the gyrokinetic Poisson equation is computed from the number density deposition at $n+1$. We require a self-consistent state between the particles and fields at each timestep, which can be expressed in terms of low-dimensional (compared to the degrees of freedom in the particle system) residuals by using the spatially discretized versions of \refeqs{eq:ampere}{eq:gkpoisson}. We define the residuals by
\begin{align}
R_A(\phi^{n+1},A_\pll^{n+1/2}) & \equiv -\nabla_\perp^2 A_\pll^{n+1/2} - \mu_0 \sum_{s=i,e} \langle \delta j_{\pll s} \rangle^{n+1/2} \nonumber \\
R_\phi(\phi^{n+1},A_\pll^{n+1/2}) & \equiv - \nabla \cdot \frac{m_i n_{0i}}{B^2} \nabla_\perp \phi^{n+1} - \sum_{s=i,e} q_s \langle \delta n_s \rangle^{n+1}. 
\end{align}
For given $\phi^{n+1}$ and $A_\pll^{n+1/2}$, evaluation of the residuals involves: pushing the marker particles from step $n$ to $n+1$, as described in the previous subsection; depositing $\langle \delta j_{\pll s} \rangle^{n+1/2}$ and $\langle \delta n_s \rangle^{n+1}$ to the mesh; and evaluating the elliptic operators in Amp\`{e}re's law and the gyrokinetic Poisson equation. Self-consistency is expressed as
\begin{align}
R_A(\phi^{n+1},A_\pll^{n+1/2}) &= 0 \nonumber \\
R_\phi(\phi^{n+1},A_\pll^{n+1/2}) &= 0, \label{eq:discpots}
\end{align}
which represents a nonlinear system of equations for $\phi^{n+1}$ and $A_\pll^{n+1/2}$. We note that both particle and field quantities are coupled in \refeq{eq:discpots}. The moments $\langle \delta j_{\pll s} \rangle^{n+1/2}$ and $\langle \delta n_s \rangle^{n+1}$ are implicitly functions of $\phi^{n+1}$ and $A_\pll^{n+1/2}$, where the dependence comes in through the particle equations of motion \cite{GChen2011}. Furthermore, given $\phi^{n+1}$ and $A_\pll^{n+1/2}$ that satisfy \refeq{eq:discpots}, the consistent particle states follow directly by pushing the particles with the resulting fields.

\subsection{Iterative Solution Method}

Our iterative solution method involves wrapping Anderson mixing \cite{Anderson1965} around a preconditioned Picard iteration scheme. Here, we give a brief overview of the preconditioned Picard iteration scheme. We define a state vector for the potentials $\vx = [\phi^{n+1},A_\pll^{n+1/2}]^T$ and a residual vector $\vr = [R_\phi,R_A]^T$. In this notation, \refeq{eq:discpots} can be stated as $\vr(\vx) = 0$, and the preconditioned Picard iteration scheme can be written as
\begin{align}
\sP \Delta \vx^k &= \vr(\vx^k) \label{eq:picard1} \\
\vx^{k+1} &= \vx^k + \Delta \vx^k \nonumber,
\end{align}
where $\sP$ is an invertible operator whose inverse maps the residual vector to a correction vector $\Delta \vx$ and $k$ is the iteration index. We refer to the operator $\sP$ as the preconditioner and note that the consistency of the iterative method does not depend on the particular form of $\sP$. In other words, any invertible preconditioner used in this scheme will produce a correct solution such that $\vr(\vx) = 0$ provided the scheme converges. The choice of preconditioner will, however, determine if (and at what rate) the scheme converges. As mentioned in the previous subsection, each evaluation of $\vr$ on the right hand side of \refeq{eq:picard1} involves a particle push and deposition to compute updated velocity moments. One preconditioned Picard iteration from $k$ to $k+1$ therefore follows these steps
\begin{enumerate}
    \item Push particles using the fields computed from the potentials in $\vx^k$.
    \item Deposit $\langle \delta j_{\pll s} \rangle$ and $\langle \delta n_s \rangle$ to compute the residuals in $\vr^k$.
    \item Solve $\sP \Delta \vx^k = \vr(\vx^k)$ to get the corrections to the potentials $\Delta \vx^k$.
    \item Update the potentials by $\vx^{k+1} = \vx^k + \Delta \vx^k$.
\end{enumerate}

\subsection{Fluid-Based Preconditioner}
\par Similarly to what has been reported elsewhere for 6D electromagnetic PIC \cite{GChen2014,GChen2014JCP,GChen2015}, we choose a preconditioner based on a simplified set of implicitly discretized fluid equations that are linearized with respect to changes in $\phi^{n+1}$ and $A_\pll^{n+1/2}$. Our starting point is to consider the continuity and momentum equations for electrons, keeping only the terms relevant to dynamics parallel to the background magnetic field. This greatly simplifies the equations we will need to solve when applying $\sP^{-1}$ to the residuals, yet still captures the terms responsible for the fastest timescales in the system. Since ions evolve on a much slower timescale than electrons, we neglect them entirely in the preconditioner. The starting electron fluid equations are
\begin{align}
\pd{}{t} \delta n_e &- \frac{1}{e} B \nabla_\pll B^{-1} \delta j_{\pll e} = 0 \\
\pd{}{t} \delta j_{\pll e} &+ \frac{e^2 n_0}{m_e} \left( \nabla_\pll \phi + \pd{}{t} A_\pll \right) - \frac{e T_e}{m_e} B \nabla_\pll B^{-1} \delta n_e = 0. \label{eq:electron_momentum}
\end{align}
Next, we consider an implicit discretization of these equations which defines quantities at discrete time locations consistent with the quantities in the PIC system. We have
\begin{align}
\frac{\delta n_e^{n+1}}{\Delta t} &- \frac{1}{e} B \nabla_\pll B^{-1} \delta j_{\pll e}^{n+1/2} = F_c^{n} \\
\frac{2}{\Delta t} \delta j_{\pll e}^{n+1/2} &+ \frac{e^2 n_0}{m_e} \left(\frac{1}{2} \nabla_\pll \phi^{n+1} + \frac{2}{\Delta t} A_\pll^{n+1/2} \right) - \frac{e T_e}{m_e} B \nabla_\pll B^{-1} \frac{1}{2} \delta n_e^{n+1} = F_M^n,
\end{align}
where the right hand sides of these equations can be determined from information known at the previous timestep. Finally, if we consider the linear responses of the fluid moments due to small perturbations in $\phi^{n+1}$ and $A_\pll^{n+1/2}$, we have :
\begin{align}
\Delta n_e &- \frac{\Delta t}{e} B \nabla_\pll B^{-1} \Delta j_{\pll e} = 0 \label{eq:discfluid} \\
\Delta j_{\pll e} &+ \frac{e^2 n_0}{m_e} \left( \frac{\Delta t}{4} \nabla_\pll \Delta \phi + \Delta A_\pll \right) - \frac{\Delta t}{4} \frac{e T_e}{m_e} B \nabla_\pll B^{-1} \Delta n_e = 0. \nonumber
\end{align}
In our notation, $\delta$ refers to perturbations from background quantities in the system, whereas $\Delta$ refers to perturbations due to small changes in the discrete-time potentials $\phi^{n+1}$ and $A_\pll^{n+1/2}$. A 4 $\times$ 4 block matrix $\mA$ can be written using \refeq{eq:discfluid} together with Amp\`{e}re's law and the gyrokinetic Poisson equation, which describes the linear response of the coupled moment-potential system. We write
\begin{equation}
\arraycolsep=1.4pt\def\arraystretch{2.2}
\mA = \left[
\begin{array}{cc|cc}
- \nabla \cdot \frac{m_i n_{0 i}}{B^2} \nabla_\perp & 0 & e \mM & 0	\\
0 & - \nabla_\perp^2 & 0 & - \mu_0 \mM	\\
\hline
0 & 0 & \mI & -\frac{\Delta t}{e} B \nabla_{\parallel} B^{-1} \\
\frac{\Delta t}{4} \frac{e^2 n_0}{m_e} \nabla_{\parallel} & \frac{e^2 n_0}{m_e} \mI & -\frac{\Delta t}{4} \frac{e T_e}{m_e} B \nabla_{\parallel} B^{-1} & \mI
\end{array}
\right], \label{eq:PCmatrix}
\end{equation}
where the operators in each block are $N\times N$ matrices with $N$ the number of mesh vertices, $\mI$ is the $N \times N$ identity matrix, and $\mM$ is a mass matrix for the finite element discretization used in XGC over the poloidal planes. We recall that in step 4 from the previous subsection only the potentials are updated at the end of each iteration. \refequation{eq:PCmatrix} represents an augmented system with equations involving the additional unknowns $\Delta n_e$ and $\Delta j_{\pll e}$. These additional unknowns can provide an intuitive way of expressing the system, where the top two rows of $\mA$ model the responses of the potentials to changes in the moments, and the bottom two rows model the responses of the moments to changes in the potentials. However, the actual solutions obtained for $\Delta n_e$ and $\Delta j_{\pll e}$ do not play a role in the update.
\par We use the block matrix system to solve for $\Delta \vx^k$ in \refeq{eq:picard1} in two steps:
\begin{enumerate}
\item Solve $\mA \vy = \mW \vr^k$, where $\vy = [\Delta \phi, \Delta A_\pll, \Delta n_e, \Delta j_{\pll e}]^T$ and 
\[\arraycolsep=8pt\def\arraystretch{2.2}
\mW = \left[
\begin{array}{cc}
\mI & 0 \\
0 & \mI \\
0 & 0 \\
0 & 0
\end{array}
\right].
\]
\item Restrict the solution to the potential correction vector $\Delta \vx^k$ with $\Delta \vx^k = \mW^T \vy$.
\end{enumerate}
The block matrix is therefore related to the preconditioner operator in \refeq{eq:picard1} by $\sP^{-1} = \mW^T \mA^{-1} \mW$. In our implementation, we use the suite of solvers available in PETSc \cite{petsc-user-ref,petsc-efficient,lidemmel03} to invert $\mA$. Some initial convergence results for the iterative scheme applied to the preconditioned implicit PIC equations are given in Sec.~\ref{sec:performance}. However, detailed studies of convergence rates across different parameter regimes, modifications to the fluid-based preconditioner for improving the convergence rates, and methods for efficiently inverting $\mA$ will be the subject of a separate paper.


\section{Verification Tests} \label{sec:verification}

Our verification tests are based on cases previously considered in \cite{Ma2018}. In this section, we describe the problem set up and present results for two test cases. The first is shear Alfv\'{e}n wave (SAW) propagation in a periodic cylindrical system and the second is a toroidal case demonstrating the transition from an ion temperature gradient (ITG) instability to the kinetic ballooning mode (KBM) as $\beta$ is increased past a critical transition value. The SAW results are compared to an analytical dispersion relation and the ITG-KBM transition results are compared to results obtained from the GEM and GENE codes, in addition to XGC results using an explicit timestepping method with the mixed variables/pullback transformation scheme \cite{Cole2021}.

\subsection{Magnetic Geometry Model}

For the toroidal system, we again consider the cylindrical coordinate system $(R,Z,\varphi)$ from Sec.~\ref{sec:implicit_timedisc}. In addition, we can define a simple toroidal coordinate system by expressing $R$ and $Z$ in terms of the $R$ coordinate at the magnetic axis $R_0$, a minor radius coordinate $r$, and a poloidal angle coordinate $\theta$ as
\begin{align}
R &= R_0 + r\cos{\theta} \\
Z &= r \sin{\theta}.
\end{align}
The background magnetic field model used is from \cite{Lapillonne2009} and can be expressed in the cylindrical coordinates as
\begin{align}
\vB = \frac{\h{\varphi} \times \nabla \Psi}{R} + \frac{B_0 R_0}{R} \h{\varphi}, \label{eq:Bfield}
\end{align}
where $\Psi$ is a poloidal flux function. An analytical form is taken for $\Psi$ depending on $r$ only as
\begin{align}
\Psi(r) = B_0 \int_0^r \frac{r' dr'}{q(r')\sqrt{1-\left(r'/R_0\right)^2}}, \label{eq:psi}
\end{align}
where $q(r)$ is the safety factor. Hence, the magnetic equilibrium is determined once we specify the parameters $B_0$ and $R_0$ and the function $q(r)$. We take $q(r)$ to be a quadratic polynomial in $r/a$, where $a$ is the minor radius
\begin{align}
q(r) = q_0 + q_1 \left(\frac{r}{a}\right) + q_2 \left( \frac{r}{a} \right)^2.
\end{align}
Hence $q(r)$ is determined by the parameters $q_0$, $q_1$, and $q_2$ in addition to the minor radius $a$. To adapt this model to the periodic cylindrical system used in the SAW benchmark, \refeq{eq:Bfield} is modified by replacing $R$ in the denominator of both terms with the constant $R_0$. The toroidal angle $\varphi$ then serves as an axial coordinate and $(R,Z)$ are coordinates within planes perpendicular to the cylindrical axis. The distance along the cylindrical axis is parameterized by $R_0 \varphi$, and the system remains $2 \pi$-periodic in $\varphi$. In this way, we can interpret the cylindrical system as a ``straightened" toroidal system.

\subsection{Model for Density and Temperature Profiles}

To describe density and temperature profiles, we first define a normalized logarithmic gradient. Letting $A$ represent either density or temperature, the normalized logarithmic gradient of $A$ is defined as
\begin{align}
\kappa_A = -R_0 \frac{d}{dr} \ln{(A)},
\end{align}
and the profile of $A$ is then
\begin{align}
A(r) = A(r_0) \exp{\left[ - \int_{r_0}^r \frac{\kappa_A(r')}{R_0} dr' \right]},
\end{align}
where $r_0$ is a reference value of $r$. We choose an analytical form for $\kappa_A$ as
\begin{align}
\kappa_A(r) = \kappa_A(r_0) \exp{\left[-\left( \frac{r-r_0}{w_A a} \right)^6 \right]}.
\end{align}
Hence profiles are specified by the parameters $r_0$, $R_0$, $a$, $w_A$, $\kappa_A(r_0)$, and $A(r_0)$.

\subsection{Shear Alfv\'{e}n Wave} \label{sec:verification_SAW}

We simulate SAW propagation in a periodic cylindrical system with uniform temperature, density, and safety factor profiles. For simplicity, ions are modeled by a uniform background density $n_0$ in addition to the ion polarization density in the gyrokinetic Poisson equation Eq.~(\ref{eq:gkpoisson}). Electrons are kinetic. Considering only the polarization response for the ions is sufficient for providing a minimal model supporting Alfv\'{e}n wave propagation, and similar models have been considered in the past, for example in Refs. \cite{Goertz1979,Hui1992,Nishioka2021}. Physically, the ion polarization response provides a cross-field current to counter the parallel electron current in a bid to maintain quasi-neutrality. In this benchmark, deuterium ions are considered along with electrons at twice the realistic mass, giving a mass ratio of  $\frac{m_i}{m_e} = 1836$. The relevant fixed parameters for this test case are given in Table \ref{tab:alfven}.
{\renewcommand{\arraystretch}{1.5}
\setlength{\tabcolsep}{15pt}
\begin{table}[!htb]
      \centering
        \begin{tabular}{ccccc}
        \toprule
        $R_0$ & $a$ & $B_0$ & $q_0$ & $T(r_0)$ \\ \midrule
        50 m & 0.5 m & 0.228 T & 2.0 & 2.0 keV \\ \bottomrule
        \end{tabular}
        \caption{Parameters for shear Alfv\'{e}n wave test case}
        \label{tab:alfven}
\end{table}
}
\par
\begin{figure}[ht!]
\begin{center}
\includegraphics[scale=.5,clip=true,trim= 0.02in .05in 0.02in .02in]{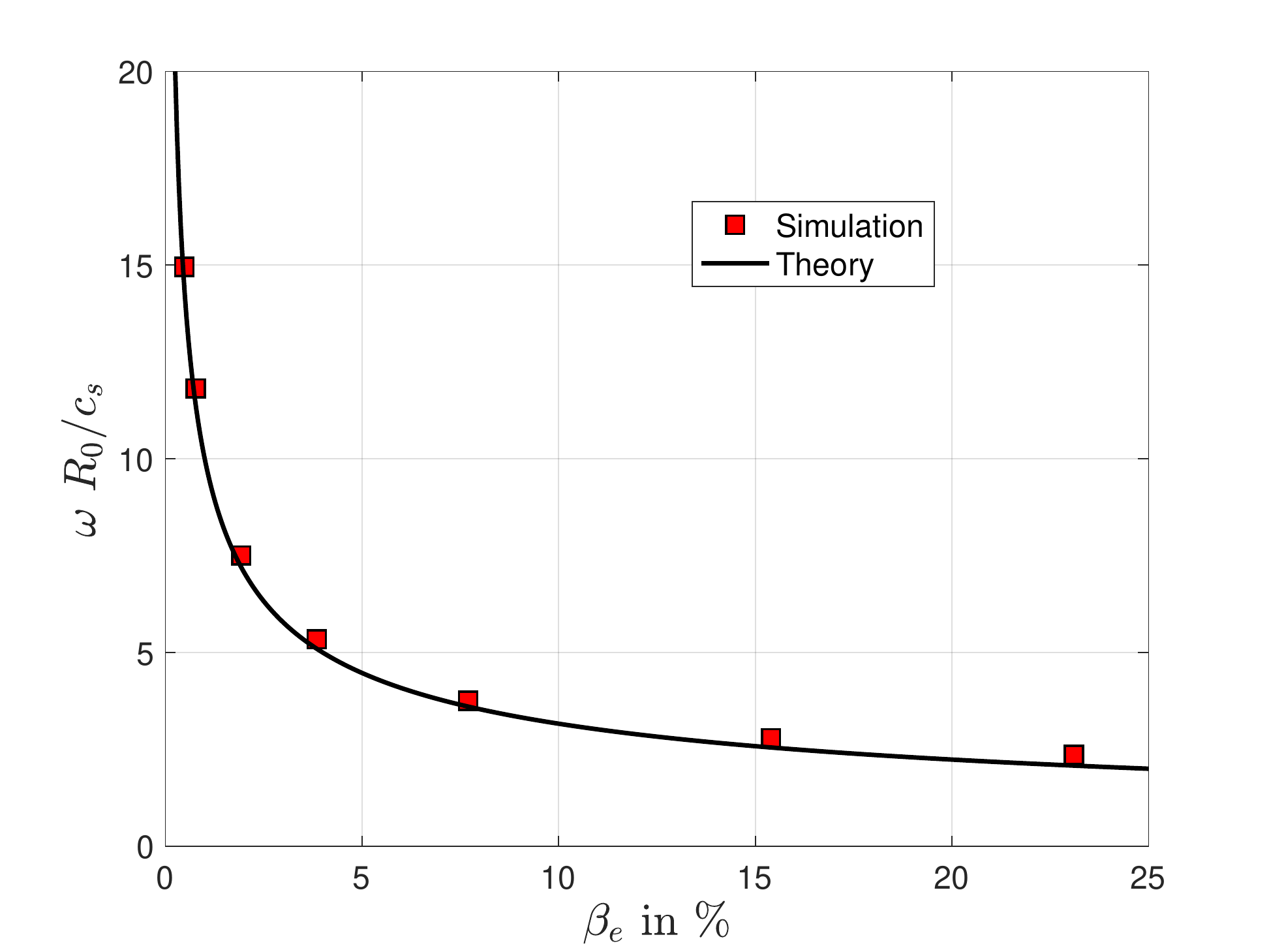}
\end{center}
\caption[]{Real frequencies vs. $\beta_e$ for the SAW test case. Results from simulations are compared to an analytical dispersion relation.}
\label{fig:alfven}
\end{figure}

 In \reffig{fig:alfven}, we scan over $\beta_e = \frac{\mu_0 n_0 T_e}{B^2}$, which we vary by choosing different values of density ranging from $6.25 \times 10^{17} \ \mathrm{m}^{-3}$ to $3.0 \times 10^{19} \ \mathrm{m}^{-3}$. Simulations are initialized with a sinusoidal electron density perturbation of the form:
\begin{align}
\delta n_e = A(r) e^{i(n \varphi + m \theta)} + \mathrm{c.c.}, \label{eq:perturbation}
\end{align}
where $A$ is a radial envelope function and c.c. denotes the complex conjugate. We note that $\beta_e$ scans of SAW propagation have long been used to verify electromagnetic gyrokinetic PIC codes. For example, \cite{Hatzky2007} and \cite{Chen2003} consider similar tests in shearless slab geometry. To shed light on potential cancellation issues, we consider a long wavelength large $\beta$ regime. We initialize our simulations taking $(n,m) = (2,2)$. Furthermore, a Fourier filter is used to keep only the $n=2$ mode. A timestep size of $\Delta t = 5.6 \times 10^{-3} R_0 / c_s$ is used, where $c_s$ is the sound speed given by $c_s = \sqrt{T_e / m_i} = 3.1 \times 10^5 \ \mathrm{m/s}$. 
\par The mesh in XGC consists of a number of identical $R-Z$ planes equally spaced in $\varphi$, where within each $R-Z$ plane, an unstructured triangular mesh is employed. For this case, 9038 mesh vertices are used within each $R-Z$ plane, corresponding to a spatial resolution of $\Delta R \approx \Delta Z \approx 9.3 \times 10^{-3} \ \mathrm{m}$, which is roughly $0.3$ times the size of the ion-sound gyroradius $\rho_s = c_s / \Omega_i = 2.8 \times 10^{-2} \ \mathrm{m}$. Between the $R-Z$ planes, a resolution of $R_0 \Delta \varphi \approx 19.5 \ \mathrm{m}$ is used. Finally, we take approximately 50 particles per mesh vertex. We compare the measured real frequencies from the simulations to a simple analytic dispersion relation. In \ref{app:AppendixB}, we present a dispersion analysis for the SAW resulting in
\begin{align}
\omega = \pm \frac{c_s}{\sqrt{\beta_e}} k_\pll. \label{eq:saw_dispersion}
\end{align}
We note that several simplifying assumptions have been made in the derivation of \refeq{eq:saw_dispersion}. For example, we have neglected kinetic effects, effects due to nonuniformities in $\vB$, and finite $k_\perp$ effects. For $r/R_0 \ll 1$, with the magnetic field described in \refeq{eq:Bfield} and \refeq{eq:psi} and perturbations of the form \refeq{eq:perturbation}, we have $k_\pll \approx (n - m/q)/R_0$. \reffigure{fig:alfven} shows excellent agreement when comparing simulation results to this approximate dispersion relation, demonstrating the absence of cancellation errors in our implementation.

\begin{figure}[ht!]
\begin{center}
\includegraphics[scale=.4,clip=true,trim= 0.02in .75in 0.02in .75in]{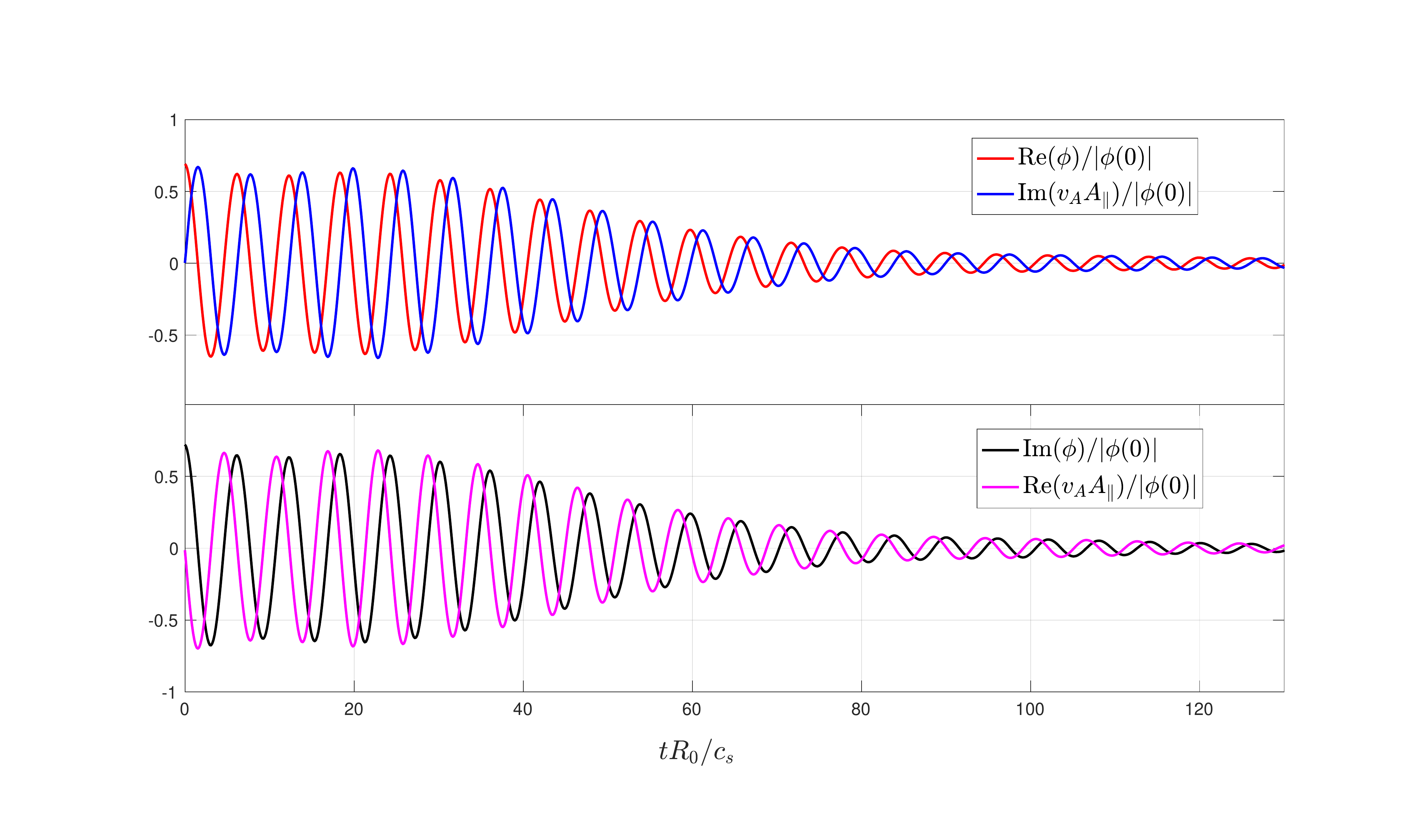}
\end{center}
\caption[]{Time histories of the real and imaginary parts of the complex amplitudes of $A_\pll$ and $\phi$ for the $(n,m) = (2,2)$ mode at $\beta_e = 3.85 \% $.}
\label{fig:alfvenhist}
\end{figure}

 In addition to comparing real frequencies, we also show the time history of the real and imaginary parts of the $(n,m) = (2,2)$ mode amplitude of $A_\pll$ and $\phi$ for the simulation at $\beta_e = 3.85\% $ in \reffig{fig:alfvenhist}. A $90^\circ$ phase shift is observed between $\mathrm{Re}(\phi)$ and $\mathrm{Im}(A_\pll)$ and between $\mathrm{Im}(\phi)$ and $\mathrm{Re}(A_\pll)$. This phase relation follows from  \refeq{eq:ampere}, \refeq{eq:gkpoisson}, and the electron continuity equation as is shown in \ref{app:AppendixA}.

\subsection{ITG-KBM Transition Benchmark}

Next, we consider a cyclone-like \cite{Dimits2000} test case in toroidal geometry similar to the case considered in \cite{Gorler2016}, demonstrating the ITG-KBM transition. ITG represents ion temperature-gradient driven modes, and KBM represents kinetic ballooning modes. In this study, we compare results from our fully implicit electromagenetic algorithm in XGC to results obtained from GEM, GENE, and explicit XGC. The GEM and GENE codes are both global $\delta f$ gyrokinetic codes using field-line-following coordinates. GEM is a particle-in-cell code, and GENE is Eulerian using a spectral method in the binormal direction. The electromagnetic capability in GEM is based on the $p_\pll$-formalism, and a modified mass matrix is used in Amp\`{e}re's law to mitigate accuracy problems at high $\beta$ from the cancellation problem \cite{Chen2007}. In GENE, $v_\pll$ is used as a coordinate for electromagnetic simulations, and a transformation of the perturbed distribution function involving $A_\pll$ is performed to eliminate the time derivative in the inductive component of the electric field \cite{Gorler2011}. GEM and GENE both use forms of the polarization density in the gyrokinetic Poisson equation that are valid for arbitrary wavelengths. In this study, on the other hand, the implicit and explicit electromagnetic XGC have the short wavelength Pad\'{e} correction term turned off, as shown in \refeq{eq:gkpoisson}.  This may cause some discrepancy for short wavelength modes, which has been shown, however, to be insignificant in the recent $n=19$ benchmarking exercise between GENE and explicit electromagnetic XGC in the G\"{o}rler benchmarking plasma \cite{Cole2021}. Our benchmarking is at the toroidal mode number $n=6$. More detailed descriptions of GEM and GENE can be found in the literature including Refs. \cite{Gorler2011PoP,Gorler2011,Gorler2009,Jenko2000,Lapillonne2010} for GENE and Refs. \cite{Chen2001,Chen2003,Chen2007} for GEM. For further comparison, we mention that XGC uses unstructured triangular meshes in cylindrical coordinates. In addition, the typical mode of operation in XGC is the total-$f$ method described in \cite{Ku2016,Ku2018}. However, in this paper XGC is run in standard $\delta f$ mode for a proper comparison to GEM and GENE.
\par In this benchmark problem, we take a reduced magnetic field strength compared to the case considered in \cite{Gorler2016}. The motivation for this choice is to reduce the resolution requirements, allowing this benchmark problem to be run at a lower computational cost. In \cite{Gorler2016}, the size parameter $\rho^*$, defined as the ratio of ion gyroradius to the minor radius, was approximately $1/180$; here, $\rho^* \approx 1/50$. Since the perpendicular resolution requirement is set by the size of the ion gyroradius, the number of mesh nodes required for a well-resolved simulation of this benchmark problem is roughly a factor of $(180/50)^2 \approx 13.0$ smaller than that required for the case considered in \cite{Gorler2016}. Both ions and electrons are treated kinetically. We take hydrogen ions and electrons with realistic mass, giving a mass ratio of $\frac{m_i}{m_e} = 1836$, and we take $T \equiv T_i = T_e$. Furthermore, only the $n=6$ mode is kept in the simulations. We note that the ITG-KBM transition case in \cite{Gorler2016} used $n=19$. Fixed parameters for defining the magnetic geometry and profiles are given in Table \ref{tab:itgkbm}.
{\renewcommand{\arraystretch}{1.5}
\setlength{\tabcolsep}{6.5pt}
\begin{table}[!htb]
      \centering
        \begin{tabular}{llllllllllll}
        \toprule
        $R_0$ & $a$ & $r_0$ & $B_0$ & $q_0$ & $q_1$ & $q_2$ & $T(r_0)$ & $\kappa_T(r_0)$ & $w_T(r_0)$ & $\kappa_n(r_0)$ & $w_n(r_0)$ \\ \midrule
        2.8 m & 1.0 m & 0.5 m & 0.236 T & 0.86 & -0.16 & 2.52 & 2.14 keV & 6.92 & 0.25 & 2.22 & 0.25 \\ \bottomrule
        \end{tabular}
        \caption{Parameters for ITG-KBM test case}
        \label{tab:itgkbm}
\end{table}
}
\par
\begin{figure}[ht!]
\begin{center}
\includegraphics[scale=.45,clip=true,trim= 0.05in .05in 0.05in .05in]{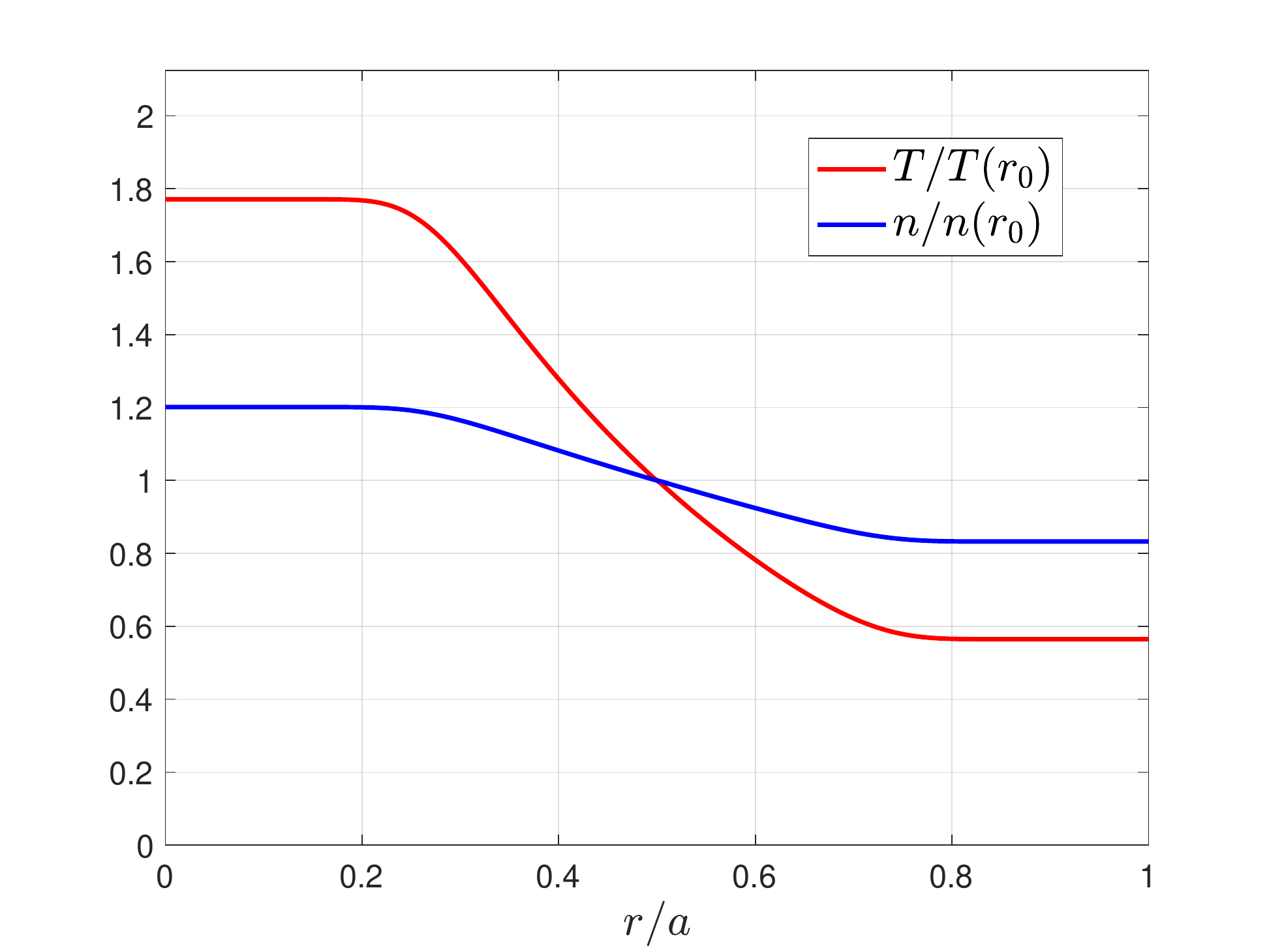}\includegraphics[scale=.45,clip=true,trim= 0.05in .05in 0.05in .05in]{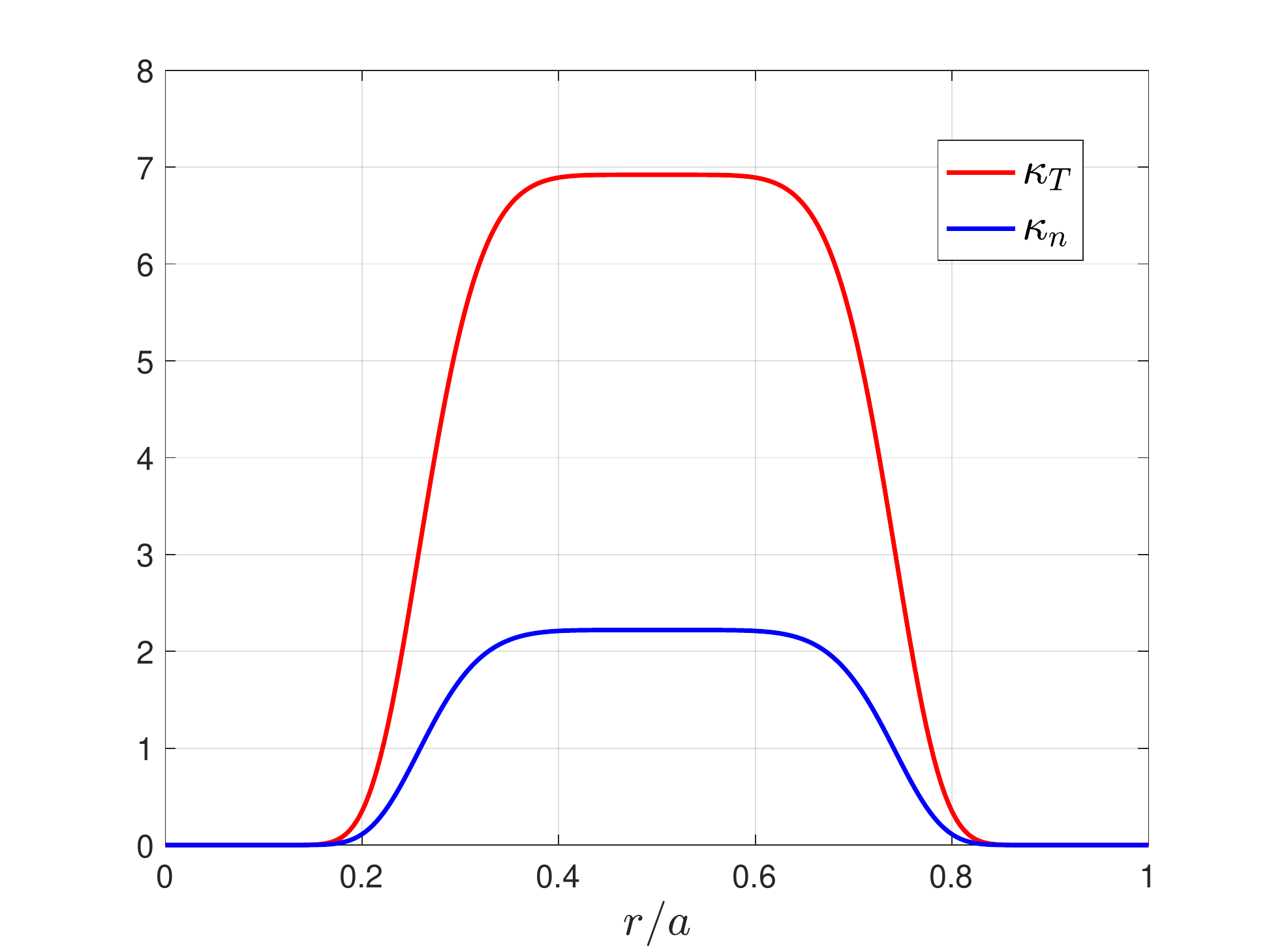}
\end{center}
\caption[]{Normalized density and temperature profiles (left) and normalized logarithmic gradients (right)}
\label{fig:profiles}
\end{figure}

\begin{figure}[ht!]
\begin{center}
\includegraphics[scale=.45,clip=true,trim= 0.05in .05in 0.05in .05in]{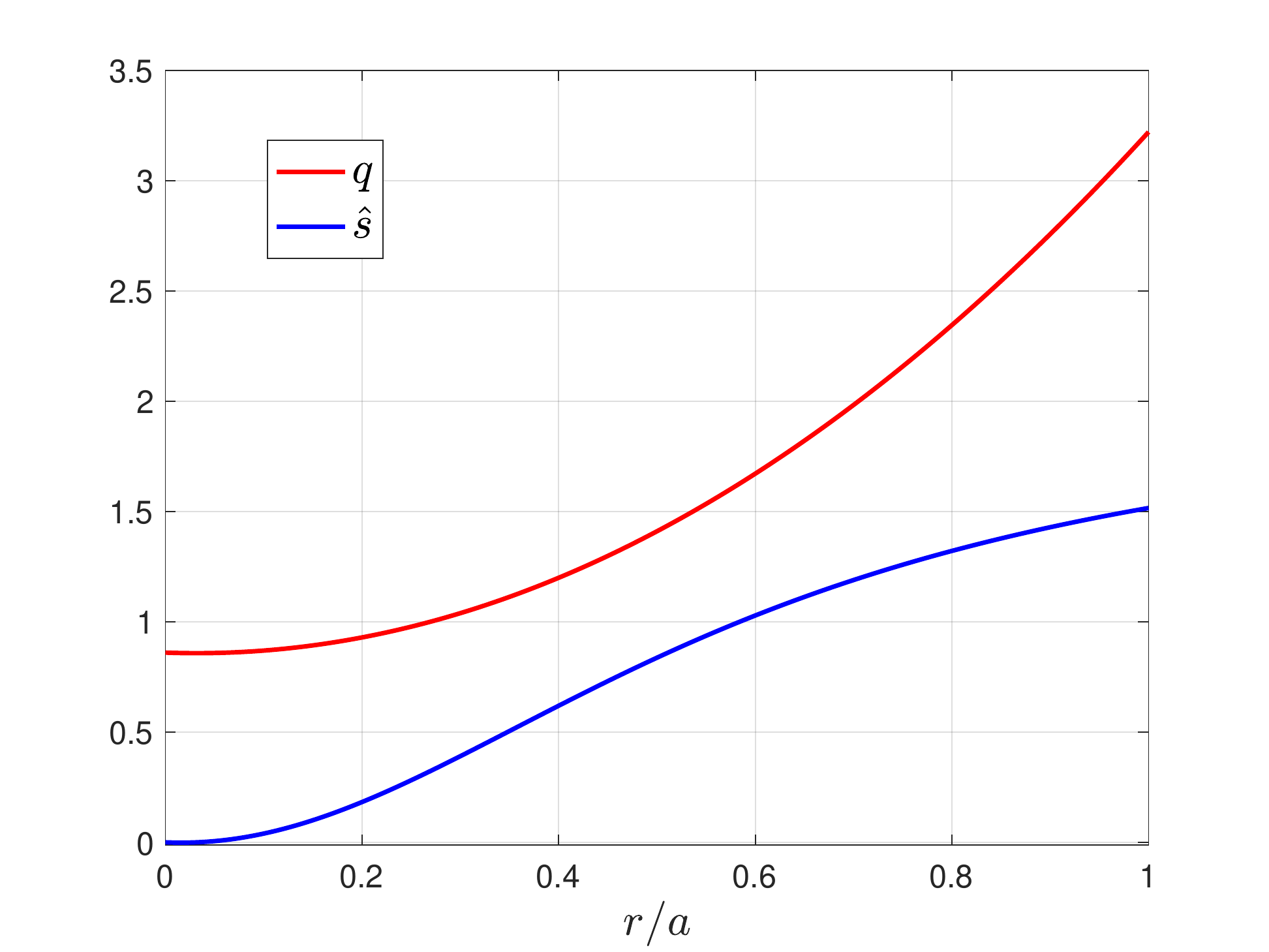}
\end{center}
\caption[]{Safety factor profile (red) and magnetic shear $\h{s} = \frac{r q'}{q}$ profile (blue)}
\label{fig:qprofile}
\end{figure}

 In \reffig{fig:profiles}, we show the density and temperature profiles normalized by their values at $r_0$ on the left and the normalized logarithmic gradients of density and temperature on the right. Compared to the profiles given in Figure 2 of \cite{Gorler2016}, the values of $\kappa_T$ and $\kappa_n$ are nearly identical at $r_0$; however, the $\kappa$ profiles considered here are much flatter in the center and nearly zero at the ends. This results in flat density and temperature profiles near the magnetic axis and the outer radial boundary. More localized gradients are used in this benchmark problem since this can help reduce effects at the boundaries of the simulation domain, which can be more significant in problems with larger $\rho^*$.
In \reffig{fig:qprofile}, the safety factor and magnetic shear profiles are shown. Note that these are identical to the ones used in \cite{Gorler2016}.
\par
\begin{figure}[ht!]
\begin{center}
\includegraphics[scale=.45,clip=true,trim= 0.05in .05in 0.05in .05in]{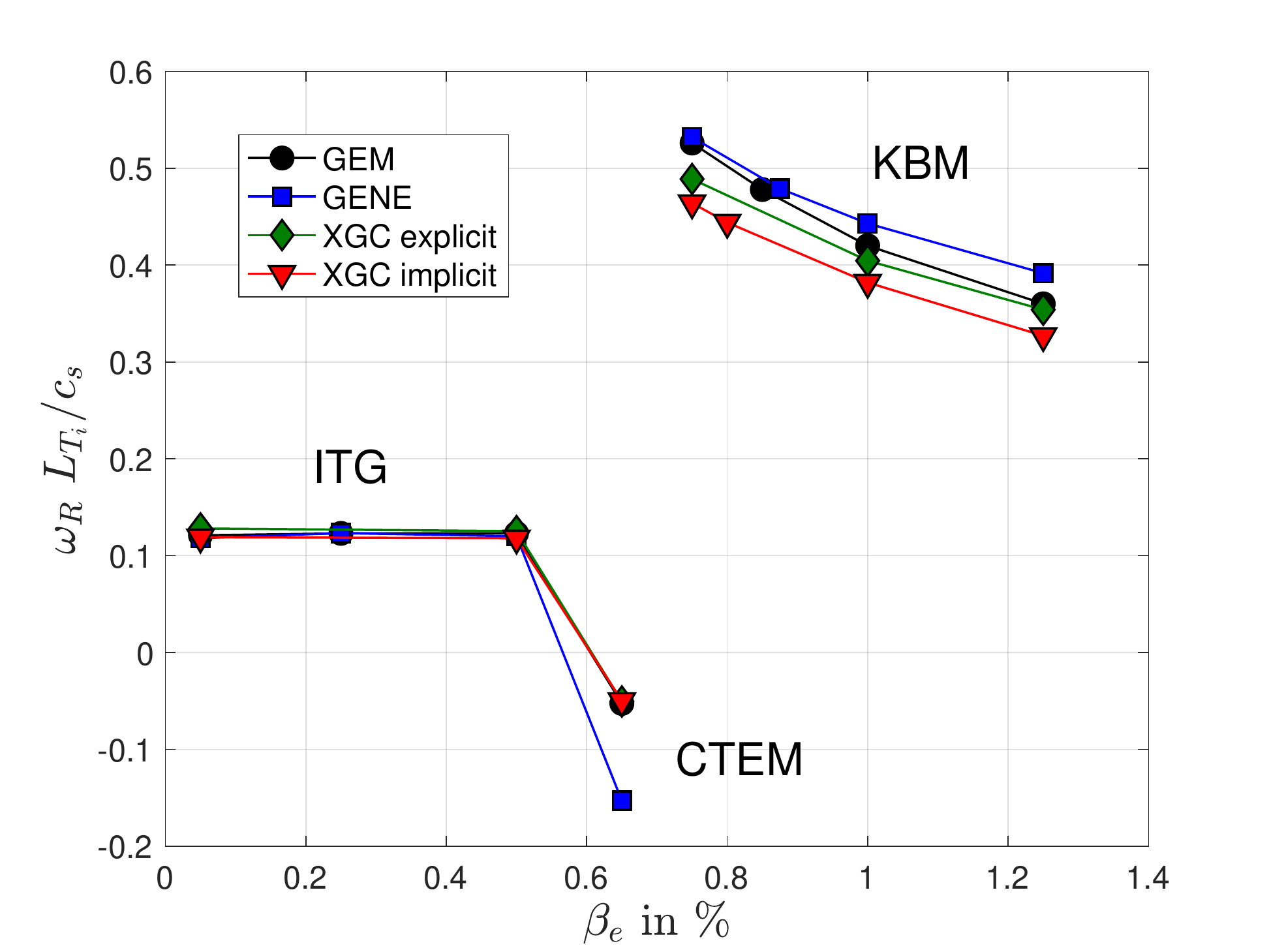}\includegraphics[scale=.45,clip=true,trim= 0.05in .05in 0.05in .05in]{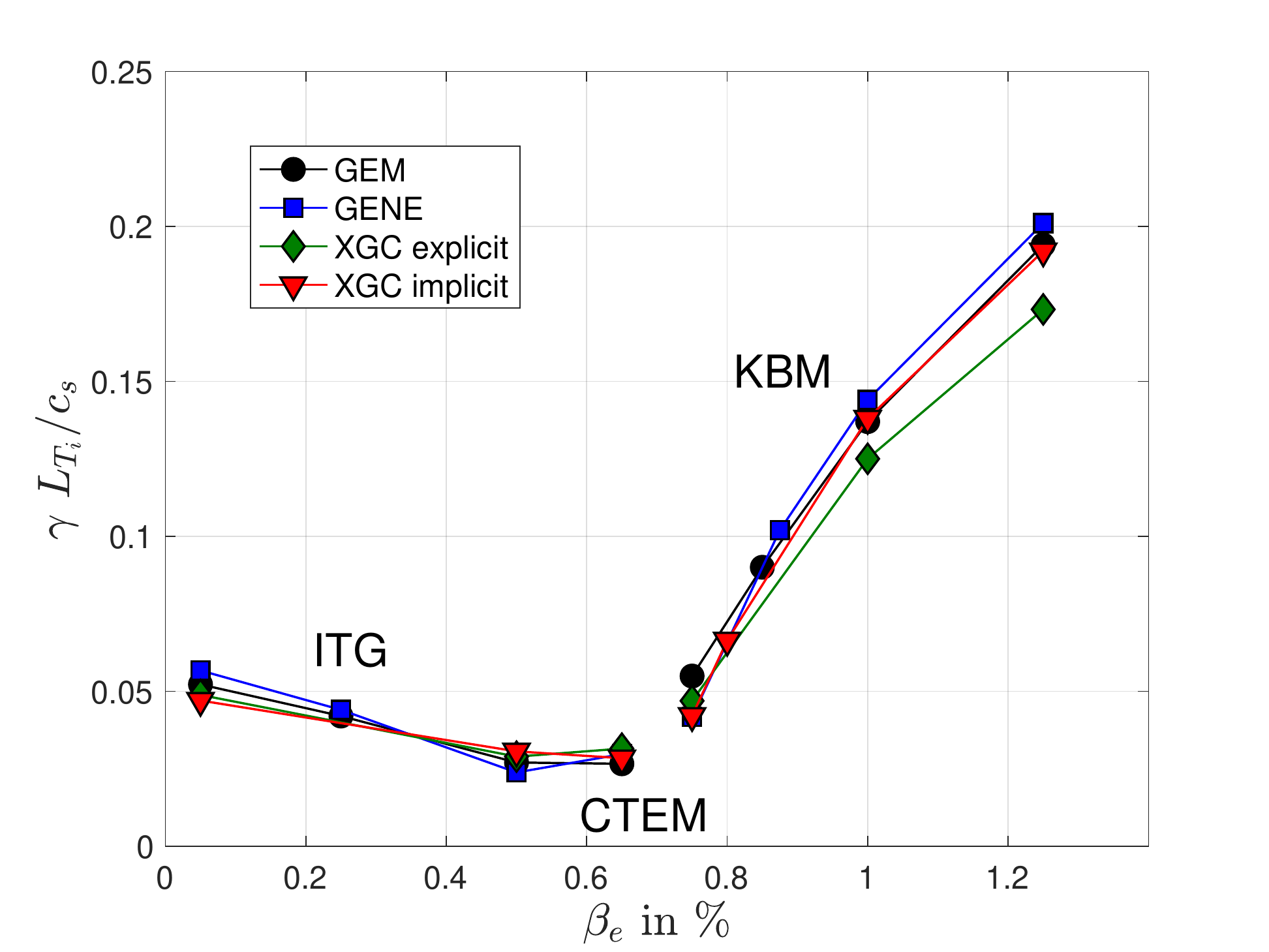}
\end{center}
\caption[]{Real frequencies (left) and growth rates (right) obtained from simulations across the various codes/implementations vs. $\beta_e$.}
\label{fig:itgkbm}
\end{figure}

 We again scan over $\beta_e$ by varying the density value of density at $r_0$ from $6.5 \times 10^{16} \ \mathrm{m}^{-3}$ to $1.625 \times 10^{18} \ \mathrm{m}^{-3}$. In \reffig{fig:itgkbm}, we compare measured real frequencies and growth rates obtained from the implicit version of XGC to those obtained in GEM, GENE, and explicit XGC. The real frequencies are given in table \ref{tab:realfreq} and the growth rates in table \ref{tab:growthrate} for the cases that were simulated by all four codes/algorithms. Good agreement is observed with each code/algorithm finding the critical value of $\beta_e$ to be between $0.65 \%$ and $0.75 \%$. Given the significant differences that exist between the algorithms and formulations used, we consider the overall agreement in real frequencies and growth rates for this benchmark case to be very good. We note that all four comparisons found a collisionless trapped electron mode (CTEM) to be present at $\beta_e = 0.65 \%$, characterized by a change in the direction of propagation from the ion diamagnetic direction to the electron diamagnetic direction. We note that there is some disagreement between GENE and the other codes in the real frequency for the CTEM case. A possible cause for this disagreement may be the form of the curvature drift that was used in GENE for this benchmarking exercise. GENE was run using the form of the curvature drift given in \ref{app:AppendixC}, which is different than what was used in XGC and GEM. Both versions of XGC and GEM used the $\vB^*$ term in \refeq{eq:xdot}, which implicitly contains the curvature drift. When a true MHD equilibrium is considered, both forms of the curvature drift should be equivalent to first order in the gyrokinetic ordering. However, MHD equilibrium is not ensured in our simplified analytic models of the magnetic field and profiles as in the G\"{o}rler benchmarking case~\cite{Gorler2016}. Since the physics of the CTEM mode depends strongly on the magnetic curvature, it may be more sensitive to differences in this term than the ITG or KBM modes. 
{\renewcommand{\arraystretch}{1.5}
\setlength{\tabcolsep}{10pt}
\begin{table}[!htb]
      \centering
        \begin{tabular}{lcccc}
        \toprule
            $\beta_e$ in \% & \ \ \ \ GEM \ \ \ \  & \ \ \ \ GENE \ \ \ \  & XGC-Explicit & XGC-Implicit \\ \midrule
            0.05 & 0.121 & 0.118 & 0.128 & 0.119 \\ 
            0.50 & 0.123 & 0.120 & 0.125 & 0.118 \\ 
            0.65 & -0.053 & -0.153 & -0.051 & -0.050 \\ 
            0.75 & 0.526 & 0.532 & 0.489 & 0.464 \\ 
            1.00 & 0.420 & 0.443 & 0.405 & 0.382 \\ 
            1.25 & 0.360 & 0.392 & 0.354 & 0.327 \\ \bottomrule
        \end{tabular}
        \caption{Real frequencies in units of $c_s / L_{T_i}$ from the plot in \reffig{fig:itgkbm}}
        \label{tab:realfreq}
\end{table}
}
{\renewcommand{\arraystretch}{1.5}
\setlength{\tabcolsep}{10pt}
\begin{table}[!htb]
      \centering
        \begin{tabular}{lcccc}
        \toprule
            $\beta_e$ in \% & \ \ \ \ GEM \ \ \ \  & \ \ \ \ GENE \ \ \ \  & XGC-Explicit & XGC-Implicit \\ \midrule
            0.05 & 0.052 & 0.057 & 0.049 & 0.047 \\ 
            0.50 & 0.027 & 0.024 & 0.029 & 0.031 \\
            0.65 & 0.027 & 0.030 & 0.032 & 0.028 \\
            0.75 & 0.055 & 0.042 & 0.047 & 0.042 \\
            1.00 & 0.137 & 0.144 & 0.125 & 0.138 \\
            1.25 & 0.194 & 0.201 & 0.173 & 0.192 \\ \bottomrule
        \end{tabular}
        \caption{Growth rates in units of $c_s / L_{T_i}$ from the plot in \reffig{fig:itgkbm}}
        \label{tab:growthrate}
\end{table}
}
\par
\begin{figure}[ht!]
\begin{center}
\includegraphics[scale=.4,clip=true,trim= 0.05in 2.0in 0.05in 1.5in]{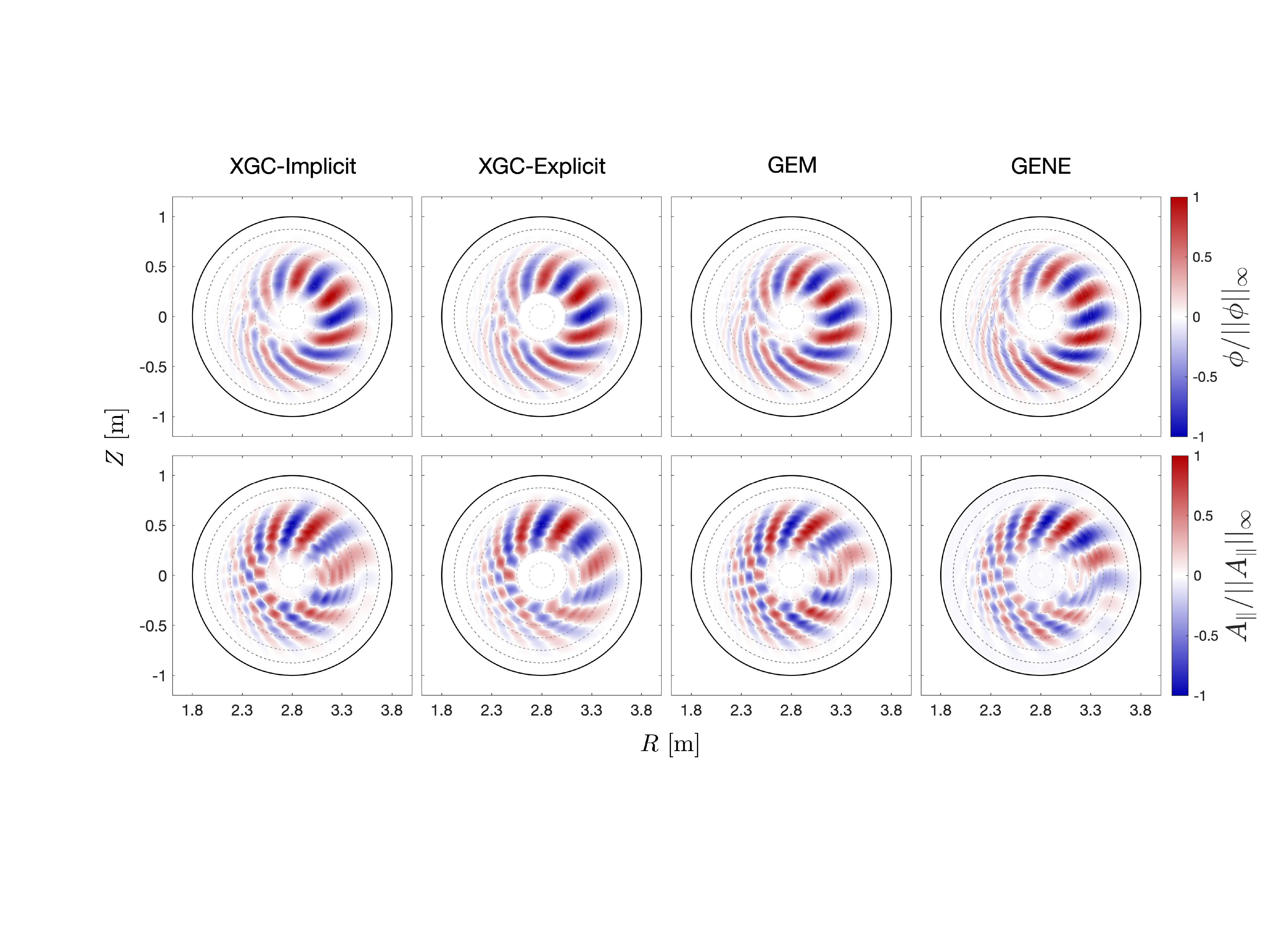}
\end{center}
\caption[]{Mode structures for $\beta_e = 0.05 \%$}
\label{fig:mslowbeta}
\end{figure}

\begin{figure}[ht!]
\begin{center}
\includegraphics[scale=.4,clip=true,trim= 0.05in 2.0in 0.05in 1.5in]{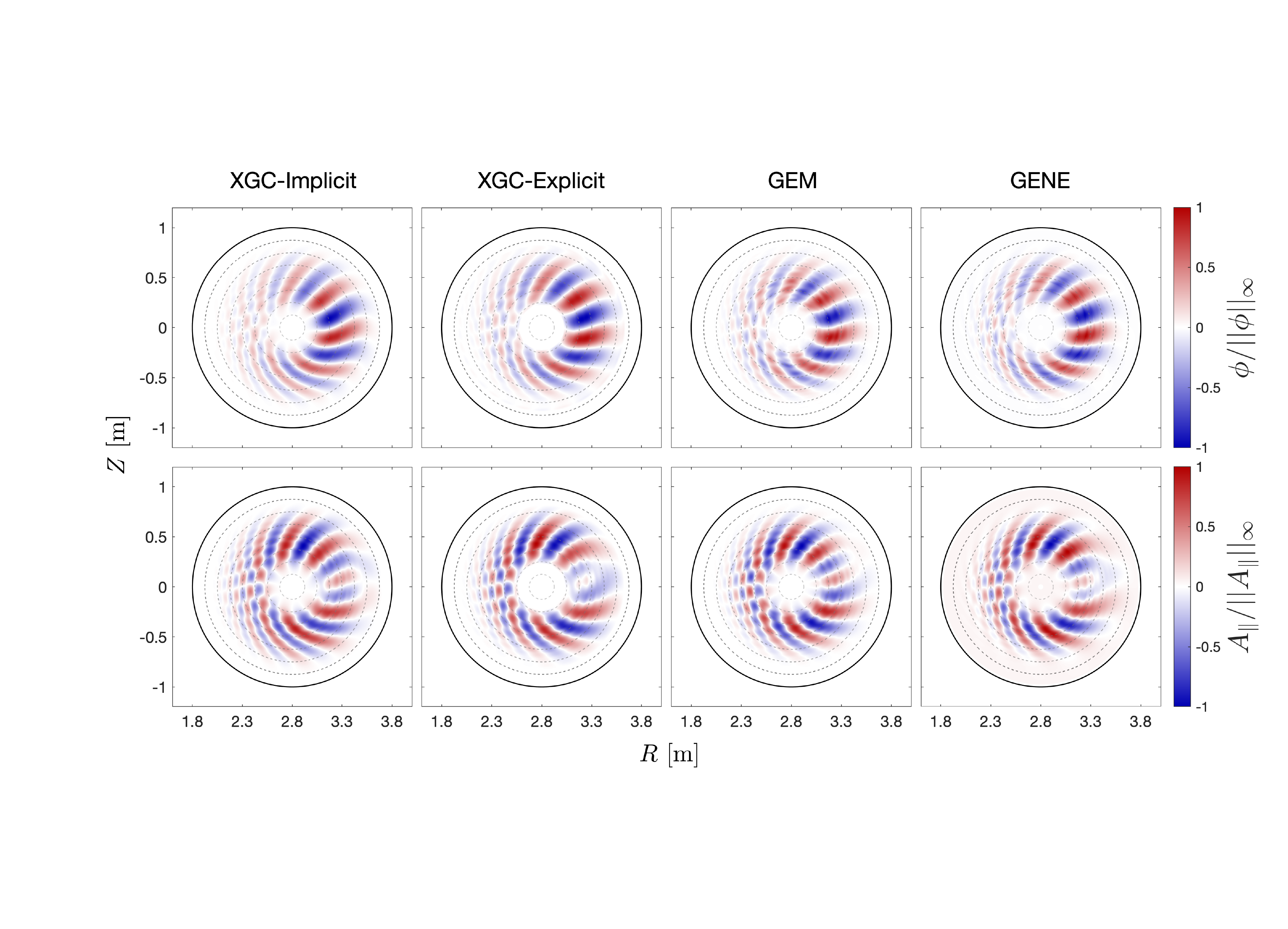}
\end{center}
\caption[]{Mode structures for $\beta_e = 1.25 \%$}
\label{fig:mshighbeta}
\end{figure}

 Finally, we show the developed mode structures for $\phi$ and $A_\pll$ in \reffig{fig:mslowbeta} for the $\beta_e = 0.05 \%$ ITG case and in \reffig{fig:mshighbeta} for the $\beta_e = 1.25 \%$ KBM case. There is good overall qualitative agreement in the mode structures between the various codes/algorithms for both cases. We note that the electrostatic potentials produced from GEM and GENE feature fine-scale radial structures near rational surfaces. The two algorithms in XGC, on the other hand, produce smoother electrostatic potentials in the radial direction. This may be due to the long-wavelength form of the ion polarization density used in \refeq{eq:gkpoisson} for the XGC algorithms.

The numerical resolutions used by each code/algorithm to obtain these results are the following. For the fully implicit version of XGC, a timestep size of $\Delta t = 4.0 \times 10^{-2} \ \frac{L_{T_i}}{c_s}$ was used, where $L_{T_i} = \frac{R_0}{\kappa_{T_i}(r_0)}$. Here, 28439 mesh vertices were used in the $R\!\!-\!\!Z$ planes, corresponding to a spatial resolution of approximately $0.5 \ \rho_i$. Along the toroidal direction, 8 $R\!\!-\!\!Z$ planes were taken within a wedge spanning $1/6$ of a full torus, and 50 particles per mesh vertex are used. The explicit version of XGC used the same mesh and toroidal resolution as was used in the implicit version. The timestep size was $\Delta t = 1.6 \times 10^{-3} \ \frac{L_{T_i}}{c_s}$ and 25 particles per mesh vertex were used. GEM used a radial resolution of $0.31 \ \rho_i$ and resolution in the binormal direction of $0.29 \ \rho_i$. Along the field line, 64 grid points were used spanning $-\pi < \theta \le \pi$. The timestep size used in GEM was $\Delta t = 5.0 \times 10^{-3} \ \frac{L_{T_i}}{c_s}$ and 16 marker ions and 32 marker electrons per cell were used. Finally, GENE used a radial resolution of $0.19 \ \rho_i$ and used a single mode in the binormal direction corresponding to a toroidal mode number of $n=6$. Along the field line, 32 grid points were used spanning $-\pi < \theta \le \pi$. In velocity space, 48 grid points were used for $v_\pll$ and 24 grid points were used for $\mu$. The timestep size used in GENE was $\Delta t = 8.8 \times 10^{-3} \ \frac{L_{T_i}}{c_s}$.

Each code used homogeneous Dirichlet boundary conditions when solving for $\phi$ and $A_\pll$. However, there was some variation in the locations of the radial boundaries. The implicit version of XGC took only an outer radial boundary at $r/a = 0.9$. Since XGC uses cylindrical coordinates, it is able to include the magnetic axis within the simulation domain. The explicit version of XGC took an outer radial boundary at $r/a = 0.98$ and an inner radial boundary at $r/a = 0.07$. In addition, filtering was applied after the field solves to set the potentials outside of the region $0.24 \le r/a \le 0.83$ to zero. GEM took an outer radial boundary at $r/a = 0.9$ and an inner radial boundary at $r/a = 0.1$. GENE took an outer radial boundary at $r/a = 0.975$ and an inner radial boundary at $r/a = 0.025$. Both versions of XGC and GEM set particle weights to zero when particles exited the simulation domain, and GENE used a buffer region outside the simulation domain with a Krook collision operator designed to damp the perturbed distribution function towards zero.
\par For completeness, we include here some timing data from each code/algorithm in simulating the $\beta_e = 1.25 \%$ case. Since the runs using the explicit version of XGC in the benchmarking study were carried out without consideration of performance, we use a more practical timestep size here for a fairer comparison. All simulations were run using 16 Haswell processor nodes on the Cori supercomputer at NESRC. Both versions of XGC and GENE ran using MPI parallelization with a total of 1024 MPI processes. GEM ran using a hybrid MPI/OpenMP parallelization with 128 MPI processes and 8 OpenMP threads per MPI process. We choose our metric to be the wallclock time required to simulate a physical time unit of $L_{T_i}/c_s$. GENE required a wallclock time of 2.4 seconds, GEM required 75.0 seconds, the implicit version of XGC required 110.3 seconds, and the explicit version of XGC required 176.6 seconds. We note that in collecting this specific timing data, the explicit version of XGC used a more practical timestep size of $1.0 \times 10^{-2} \frac{L_{T_i}}{c_s}$ and 50 particles per mesh vertex, in an attempt to improve the unnecessarily small timestep size used in the physics benchmarking exercise in the explicit version of XGC and to make the particle number the same. All other parameters are the same as those used in the benchmarking exercise presented above in this section. We caution here that the timing data comes from a single common simulation scenario in which parameters were chosen to provide a low cost physics benchmarking problem without considering computational algorithm optimization, number of subcycling steps (4 steps used in this study is at the low end), optimal hardware conditions, programing language, and practical problem size per each code. For example, optimization of the production XGC has been known to scale well for big physics problems on extreme scale computers that are equipped with high-performance accelerators. In addition, optimal choices of parameters for balancing both accuacy and efficiency are yet to be determined for the different versions of XGC. Hence, the performance numbers provided here should not be used to draw any conclusions about the efficiency of the codes/algorithms in simulating different realistic problems targeted by each codes.


\section{Performance Discussion for the Fully Implicit Scheme} \label{sec:performance}

While a detailed study of performance is beyond the scope of this paper, for completeness, we include some general discussion and outlook regarding the performance of the fully implicit scheme in addition to some initial convergence results for the SAW benchmark problem. As is the case for any simulation method, assessing performance is a complex issue requiring systematic studies over broad ranges of both physical and numerical parameters on multiple leadership class computer architectures. Here, we highlight some of the key factors contributing to the performance of the fully implicit method. We provide a rough estimate for the cost of the implicit scheme in simulating a reference unit of time $\Delta t_\mathrm{ref}$ as follows. We assume that the process cost $\mathrm{CP}_\mathrm{push}$ of pushing particles over one timestep interval $\Delta t$ is proportional to the total number of marker particles, i.e.
\begin{align}
\mathrm{CP}_\mathrm{push} = C_p N_{ppm} N_m,
\end{align}
where $N_m$ represents the mesh degrees of freedom, $N_{ppm}$ is the number of particles per mesh degree of freedom, and $C_p$ is the cost to push a single particle a time unit of $\Delta t$. Note that we assume subcycling is accounted for in the factor $C_p$, i.e., for a fixed sub-timestep size $\delta t$, $C_p$ will be proportional to the ratio of $\Delta t / \delta t$. Next, we denote the cost for applying the preconditioner, i.e. solving the linear system of equations defined by the matrix in \refeq{eq:PCmatrix}, by $\mathrm{CP}_\mathrm{pc}$ and the number of residual evaluations required per timestep for convergence of the iterative scheme by $N_\mathrm{RE}$. For the Picard scheme with Anderson mixing considered in this paper, only one residual evaluation is required per iteration, meaning $N_\mathrm{RE}$ corresponds to the number of iterations required for convergence. Our estimate for the total cost of simulating a reference unit of time is
\begin{align}
\mathrm{CP}_\mathrm{total} = \frac{\Delta t_\mathrm{ref}}{\Delta t} N_\mathrm{RE} \left( \mathrm{CP}_\mathrm{push} + \mathrm{CP}_\mathrm{pc} \right).
\end{align}
Since the cost is proportional to the number of residual evaluations $N_\mathrm{RE}$, an efficient iterative method is essential to the overall performance of the scheme. The efficiency of the iterative method relies heavily on the quality of the preconditioner. We desire a preconditioner that can effectively remove stiffness in the problem for the variety of simulation scenarios that may be encountered in practice. In addition, the cost of applying the preconditioner should not be the dominant factor in the overall cost, i.e. $\mathrm{CP}_\mathrm{pc} < \mathrm{CP}_\mathrm{push}$. For simulations requiring a large number of mesh degrees of freedom, a scalable $\sO(N_m)$ solver for handling the linear system defined by the matrix \refeq{eq:PCmatrix} becomes essential. Designing scalable solvers in this context is more challenging than for typical gyrokinetic field solves, which are decoupled in the parallel direction. The system defining the fluid-based preconditioner, on the other hand, is fully three dimensional. To address this issue, we are currently exploring the use of a Schur complement formulation of the block matrix system of \refeq{eq:PCmatrix} which is amenable to multigrid methods. This work will be reported in a future detailed performance assessment.
\par As an initial test of the performance of the iterative method, we show in Table \ref{tab:vteconverge} the number of iterations (residual evaluations) required for the SAW test case to achieve a relative tolerance of $10^{-5}$ in the preconditioned residual norm over a range of the physical parameter $\beta_e$ and the numerical parameter $\frac{v_{te} \Delta t}{R_0 \Delta \varphi}$, where $v_{te} = \sqrt{T_e/m_e}$ is the electron thermal velocity. This numerical parameter is related to the number of cell crossings for thermal electrons near the magnetic axis within the timestep size $\Delta t$. Our experience suggests that this is the most relevant parameter in determining the convergence rate of our iterative method. Our convergence criteria is
\begin{align}
\norm{\Delta \phi}_2 + v_A \norm{\Delta A_\pll}_2 < \epsilon_{\mathrm{rel}} \left( \norm{\phi}_2 + v_A \norm{A_\pll}_2 \right),
\end{align}
where $\Delta \phi$ and $\Delta A_\pll$ are the corrections to the potentials coming from the solution of the linear system defined by \refeq{eq:PCmatrix}, $\phi$ and $A_\pll$ are the values of the potentials at the current iteration, and $\epsilon_{\mathrm{rel}}$ is the relative tolerance. The reported iteration counts represent averages over 15 timesteps. 
We consider three values of $\beta_e$ for the SAW test and vary the timestep size to set the parameter $\frac{v_{te} \Delta t}{R_0 \Delta \varphi}$. Subcycling is adjusted in each case to keep $\frac{v_{te} \delta t}{R_0 \Delta \varphi}$ at the fixed value of $0.125$. We mention that a small subcycling timestep compared to the cell crossing time seems to be a necessary requirement for convergence. In the recent explicit XGC code, we subcycle the electrons 12 timesteps per ion timestep. This should be considered anyway in order for marker electrons to accurately sample the perturbed fields along their trajectories. The results from Table \ref{tab:vteconverge} show that convergence can be achieved within a reasonable number ($<10$) of residual evaluations for a wide range of $\beta_e$ and with the number of thermal electron cell crossings of order unity. When the number of electron cell crossings is pushed beyond order unity, the iterative method generally fails to converge unless further corrections are included in the preconditioner. These will be reported in future studies.
\setlength{\tabcolsep}{15pt}
\renewcommand{\arraystretch}{1.5}
\begin{table}[!htb]
  \centering
  \begin{tabular}{lccc}
    \toprule
      & \multicolumn{3}{c}{\# iterations} \\ \cmidrule{2-4}
      $\beta_e$ in \% &  $\frac{v_{te} \Delta t}{R_0 \Delta \varphi}$ = 0.25 & $\frac{v_{te} \Delta t}{R_0 \Delta \varphi}$ = 1.00 & $\frac{v_{te} \Delta t}{R_0 \Delta \varphi}$ = 4.00 \\ \cmidrule{1-4}
    0.5 & 3.9 & 5.0 & 5.5 \\
    7.7 & 5.0 & 6.2 & 7.1 \\
    23.1 & 5.9 & 7.0 & 8.3 \\ \bottomrule
  \end{tabular}
  \caption{Iterations required to achieve a relative tolerance of $10^{-5}$ in preconditioned residual norm vs $\beta_e$ and thermal electron cell crossings per timestep.}
  \label{tab:vteconverge}
\end{table}

\par In general, the cost per timestep with an implicit method will be greater than with an explicit method. The typical motivation for using an implicit method is to relax constraints on the timestep size required for numerical stability when explicit methods are applied. In our case, however, an implicit method allows us to use a form of the electromagnetic gyrokinetic equations that avoids the Amp\`{e}re cancellation problem. For problems with $\beta_e > m_e/m_i$, we have $v_{te} > v_A$, where $v_A = B/\sqrt{\mu_0 m_i n_{0i}}$ is the Alfv\'{e}n velocity. Since the convergence of our current implementation seems to require $\frac{v_{te} \Delta t}{R_0 \Delta \varphi}$ to be order unity, a large increase in the timestep size compared to an explicit method might not be feasible in this regime. The advantage instead resides in a greater robustness to the cancellation problem, and in possible efficiency gains from relaxed perpendicular spatial resolution requirements (i.e., resolving the electron skin depth is no longer needed for numerical reasons). In low density/low-$\beta$ regimes, however, there may be a significant timestep size advantage when using an implicit scheme. In this case, it is the timescale of the SAW which sets the constraint on the timestep size for explicit methods. We explore this regime in Table \ref{tab:vaconverge}, where we fix $\frac{v_{te} \Delta t}{R_0 \Delta \varphi} = 0.25$ and lower the density to increase the Alfv\'{e}n velocity (decrease $\beta_e$). Again, the iteration counts represent averages over 15 timesteps. It is demonstrated that the number of iterations required to achieve a relative tolerance of $10^{-5}$ does not increase significantly with $v_A$, even for large values of the Courant number due to the Alfv\`{e}n wave: $\frac{v_A \Delta t}{R_0 \Delta \varphi}$.
\setlength{\tabcolsep}{15pt}
\renewcommand{\arraystretch}{1.5}
\begin{table}[!htb]
  \centering
  \begin{tabular}{lcc}
    \toprule
    $\beta_e$ in \% & $\frac{v_A \Delta t}{R_0 \Delta \varphi}$ & \# iterations \\ \cmidrule{1-3}
    $2.8 \times 10^{-2}$ & 0.1 & 3.3  \\
    $2.8 \times 10^{-4}$ & 1.0 & 4.0  \\
    $2.8 \times 10^{-6}$ & 10.0 & 4.0 \\ \bottomrule
  \end{tabular}
  \caption{Iterations required to achieve a relative tolerance of $10^{-5}$ in preconditioned residual norm for fixed $\frac{v_{te} \Delta t}{R_0 \Delta \varphi}$ and increasing Alfv\'{e}n velocity (decreasing $\beta_e$).}
  \label{tab:vaconverge}
\end{table}

\par Although we have considered only the SAW benchmark problem for the convergence results in this section, in practice we have found the convergence behavior to be qualitatively similar for the ITG-KBM benchmarking problem. The inclusion of kinetic ions and background gradients in the system seems to have little effect on convergence, which is mainly determined by the fastest timescales in the problem.


\section{Conclusions} \label{sec:conclusions}

In this paper, we have verified a fully implicit $\delta f$ implementation of the $v_\pll$-formalism of electromagnetic gyrokinetics in the XGC code with two test cases - shear Alfv\'{e}n wave (SAW) propagation in cylindrical geometry and the ITG-KBM transition in toroidal geometry. The $v_\pll$-formalism uses the original form of Amp\`{e}re's law and is therefore free from the ``cancellation problem" that plagues codes based on the $p_\pll$-formalism. In addition, a fully implicit time discretization scheme can provide stable simulations for long wavelengths and high $\beta$ using the $v_\pll$-formalism despite the appearance of a time derivative in the definition of the electric field. For the SAW propagation test, the real frequencies measured from XGC simulations using the implicit electromagnetic implementation match well with an analytical dispersion relation, demonstrating the lack of a cancellation problem. The elimination of the cancellation problem with this method can be a significant advantage for long wavelength, high-$\beta$ regimes. We note that not all formulations avoiding the $p_\pll$-formalism eliminate the cancellation issue. For instance, the mixed variables/pullback transformation scheme \cite{Mishchenko2014a,Mishchenko2014b,Kleiber2016} does not use the $p_\pll$-formalism, but it still has a cancellation problem. For the ITG-KBM case, we have compared the real frequencies and growth rates obtained using the implicit electromagnetic implementation against the GEM and GENE codes in addition to an explicit electromagnetic implementation in XGC based on the mixed variables/pullback transformation scheme \cite{Cole2021}. All codes give good agreement for this case and predict the transition to occur for $\beta_e$ between $0.65 \%$ and $0.75 \%$. In addition, all codes predict a collisionless trapped electron mode to be dominant for $\beta_e = 0.65 \%$. These results strengthen our confidence in the ability of the implicit scheme to accurately solve the electromagnetic gyrokinetic equations. 

Although this paper has focused on linear verification studies for the implicit electromagnetic algorithm, verification of the fully implicit scheme for nonlinear electromagnetic plasma dynamics is an important future objective. For conventional $\delta f$ turbulence simulations in the core, where perturbation sizes are small compared to the background, electron physics parallel to the background magnetic field accounts for the main source of stiffness in the implicit equations. Since this difficulty is already present in the linear regime, we do not anticipate additional difficulties arising in the nonlinear regime. For simulations in which large perturbation sizes may develop, however, certain modifications might be necessary in the preconditioner to avoid degradation of convergence rates. 
For example, the background density and temperature may need to be updated periodically in \refeq{eq:PCmatrix} so that the preconditioner is guaranteed to be linearized against the previously updated values, or perhaps smaller timestep sizes will be required to improve the conditioning of the implicit equations. We emphasize, however, that poor approximations to the plasma dynamics in the preconditioner will only affect the performance of the scheme by yielding slower convergence rates. This will not affect the accuracy of the discrete formulation. Future work will report on further details regarding the optimization of the implicit algorithm and its performance in both linear and nonlinear regimes.

\appendix


\section{Theoretical relationship between $A_\pll$ and $\phi$} \label{app:AppendixA}

Here we derive a relationship between the complex mode amplitudes of $A_\pll$ and $\phi$ for the shear Alfv\'{e}n wave model considered in Sec.~\ref{sec:verification}. We begin by taking the partial derivative with respect to time of \refeq{eq:gkpoisson}, assuming spatially uniform background quantities and no time dependence in the ion gyrocenter density. We have
\begin{align}
- \frac{m_i n_{0i}}{B^2} \nabla_{\perp}^2 \pd{\phi}{t} \approx - e \pd{\delta n_e}{t} \label{eq:ddtpoisson}
\end{align}
Next, we take the gradient in the parallel direction of \refeq{eq:ampere}. Assuming parallel and perpendicular derivative operators commute and no contribution to the current from ion gyrocenters, we have
\begin{align}
-\frac{1}{\mu_0} \nabla_\perp^2 \nabla_\pll A_\pll \approx \nabla_\pll \delta j_{\pll e}. \label{eq:ddzampere}
\end{align}
If we assume the main contribution in the electron continuity equation to be from the parallel electron current, we have
\begin{align}
\pd{}{t} \delta n_e - \frac{1}{e} \nabla_\pll \delta j_{\pll e} \approx 0. \label{eq:econt}
\end{align}
Adding together \refeq{eq:ddtpoisson} and \refeq{eq:ddzampere}, it follows from \refeq{eq:econt} that
\begin{align}
\pd{\phi}{t} + \frac{B_0^2}{\mu_0 m_i n_{0i}} \nabla_\pll A_\pll \approx 0. \label{eq:phi_apar}
\end{align}
Finally, we take a Fourier mode ansatz along the parallel direction with wave number $k_\pll$. The complex mode amplitudes then follow
\begin{align}
\pd{\tilde{\phi}}{t} + i k_\pll v_A^2 \tilde{A_\pll} \approx 0,
\end{align}
which can be written as
\begin{align}
\pd{}{t} \mathrm{Re}(\tilde{\phi}) - k_\pll v_A^2 \mathrm{Im}(\tilde{A_\pll}) & \approx 0 \\
\pd{}{t} \mathrm{Im}(\tilde{\phi}) + k_\pll v_A^2 \mathrm{Re}(\tilde{A_\pll}) & \approx 0
\end{align}
by separating real and imaginary parts.


\section{Simple dispersion relation for the shear Alfv\'{e}n wave} \label{app:AppendixB}
A simple dispersion relation for the shear Alfv\'{e}n wave can be derived using \refeq{eq:phi_apar}, Amp\`{e}re's law \refeq{eq:ampere}, and the electron momentum equation \refeq{eq:electron_momentum} in the limit $T_e \rightarrow 0$. The electron momentum equation and Amp\`{e}re's law can be combined to give
\begin{align}
\left(1 - \frac{m_e}{e^2 n_{0} \mu_0} \nabla_\perp^2 \right) \pd{A_\pll}{t} + \nabla_\pll \phi \approx 0.
\end{align}
For long perpendicular wavelengths compared to the electron skin depth $d_e = \sqrt{m_e / e^2 n_0 \mu_0}$, the perpendicular Laplacian term can be ignored giving
\begin{align}
\pd{A_\pll}{t} \approx - \nabla_\pll \phi. \label{eq:epar_zero}
\end{align}
Taking a partial derivative with respect to time of \refeq{eq:phi_apar} and using \refeq{eq:epar_zero} gives
\begin{align}
\ppd{\phi}{t} - \frac{B_0^2}{\mu_0 m_i n_{0i}} \nabla_\pll^2 \phi \approx 0.
\end{align}
Finally, taking a plane wave ansatz gives
\begin{align}
\omega^2 = \frac{B_0^2}{\mu_0 m_i n_{0i}} k_\pll^2 = \frac{c_s^2 k_\pll^2}{\beta_e}.
\end{align}


\section{Form of curvature drift used in GENE} \label{app:AppendixC}

Here, we derive the form of the curvature drift that was used in GENE for the ITG-KBM benchmarking problem. We begin with a standard form of the curvature drift given by
\begin{align}
\vv_c = \frac{m_s}{q_s B} v_\pll^2 \h{\vb} \times \left(\h{\vb} \cdot \nabla \right) \h{\vb}.
\end{align}
Using standard vector calculus identities, we can write the second factor in the cross product as
\begin{align}
\left(\h{\vb} \cdot \nabla \right) \h{\vb} = \frac{\nabla \times \vB}{B} \times \h{\vb} + \frac{\nabla_\perp B}{B}.
\end{align}
From Amp\`{e}re's law, it follows that
\begin{align}
\vv_c = \frac{m_s}{q_s B} v_\pll^2 \h{\vb} \times \left( \mu_0 \frac{\vj \times \vB}{B^2} + \frac{\nabla B}{B} \right).
\end{align}
Finally, assuming an MHD equilibrium allows us to express the curvature drift in terms of the pressure gradient as
\begin{align}
\vv_c = \frac{m_s}{q_s B} v_\pll^2 \h{\vb} \times \left( \mu_0 \frac{\nabla P}{B^2} + \frac{\nabla B}{B} \right), \label{eq:vc_gene}
\end{align}
which is the form used in GENE for the ITG-KBM benchmarking problem considered in this paper. We note that the full form of \refeq{eq:vc_gene} is used in this study. In particular, we do not set $\nabla P$ to zero as is commonly done under the assumption of a low-pressure plasma \cite{Chen2003,Lapillonne2010}.


\section*{Acknowledgments}

This research was supported jointly by the SciDAC-4 project High-fidelity Boundary Plasma Simulation (HBPS) funded by U.S. DOE Office of Science, Office of Advanced Scientific Computing Research and Office of Fusion Energy Sciences, under Contract No. DE-AC02-
09CH11466; and the Exascale Computing Project (17-SC-20-SC), a collaborative effort of the U.S. Department of Energy Office of Science and the National Nuclear Security Administration. This research used resources of the National Energy Research Scientific Computing Center (NERSC), a U.S. Department of Energy Office of Science User Facility operated under Contract No. DE-AC02-05CH11231.


\section*{Data Availability}
Data included in the tables and figures of this paper will be made openly available in the PPPL Theory Department ARK, Ref. \cite{DataARK}. Additional data supporting the findings of this study are available from the corresponding author upon reasonable request.

\bibliography{mybibfile}

\end{document}